\documentclass{aa}  

\usepackage{pifont}
\usepackage{amsmath}
\usepackage{graphicx}
\usepackage{txfonts}
\usepackage{url}
\usepackage{subfigure}
\usepackage{longtable}
\usepackage{lscape}
\usepackage{natbib}
\usepackage[colorlinks,linkcolor=red,anchorcolor=green,citecolor=blue]{hyperref}

\newcommand{\HII}{H {\small{II}} }
\newcommand{\kms}{{\rm km~s}^{-1}}
\newcommand{\Msun} {M_{\sun}}

\newcommand{\mjyb}{{\rm mJy~beam}^{-1}}

\newcommand{\jyb}{{\rm Jy~beam}^{-1}}

\newcommand{\nhp}{\rm N_2H^+}
\newcommand{\co}{\rm C^{18}O}
\newcommand{\hcop}{\rm HCO^+}

\begin{document}

   \title{A multi-wavelength observation and investigation of six infrared dark clouds}
    \authorrunning{C.-P. Zhang et al.}
    \titlerunning{Six infrared dark clouds}

   \author{Chuan-Peng Zhang
          \inst{1,4}
          \and
          Jing-Hua Yuan\inst{1}
          \and
          Guang-Xing Li\inst{2}
          \and
          Jian-Jun Zhou\inst{3}
          \and
          Jun-Jie Wang\inst{1,4}
          }

   \institute{National Astronomical Observatories, CAS, 20A Datun Road, Chaoyang District, 100012 Beijing, China\\
   \email{cpzhang@nao.cas.cn}
     \and
    University Observatory Munich, Scheinerstrasse 1, D-81679 Munich, Germany
     \and
    Xinjiang Astronomical Observatory, CAS, 150, Science 1-street, 830011 Urumqi, China
     \and
    NAOC-TU Joint Center for Astrophysics, 850000 Lhasa, PR China
             }

   \date{Received XXX, XXX; accepted XXX, XXX}


  \abstract
   {Infrared dark clouds (IRDCs) are ubiquitous in the Milky Way, yet they play a crucial role in breeding newly-formed stars.}
   {With the aim of further understanding the dynamics, chemistry, and evolution of IRDCs, we carried out multi-wavelength observations on a small sample.}
   {We performed new observations with the IRAM 30 m and CSO 10.4 m telescopes, with tracers $\hcop$, HCN, $\nhp$, $\co$, DCO$^+$, SiO, and DCN toward six IRDCs G031.97+00.07, G033.69-00.01, G034.43+00.24, G035.39-00.33, G038.95-00.47, and G053.11+00.05. }
   {We investigated 44 cores including 37 cores reported in previous work and seven newly-identified cores. Toward the dense cores, we detected 6 DCO$^+$, and 5 DCN lines. Using pixel-by-pixel spectral energy distribution (SED) fits of the \textit{Herschel} 70 to 500 $\mu$m, we obtained dust temperature and column density distributions of the IRDCs. We found that $\nhp$ emission has a strong correlation with the dust temperature and column density distributions, while $\co$ showed the weakest correlation. It is suggested that $\nhp$ is indeed a good tracer in very dense conditions, but $\co$ is an unreliable one, as it has a relatively low critical density and is vulnerable to freezing-out onto the surface of cold dust grains. The dynamics within IRDCs are active, with infall, outflow, and collapse; the spectra are abundant especially in deuterium species.}
   {We observe many blueshifted and redshifted profiles, respectively, with $\hcop$ and $\co$ toward the same core. This case can be well explained by model ``envelope expansion with core collapse (EECC)''.}


   \maketitle
%

\section{Introduction}    
\label{sect:intro}

Ubiquitously in the Milky Way, infrared dark clouds (IRDCs) represent an important phase of interstellar medium (ISM) evolution and star formation. However, the origin of filamentary cloud structure is unclear. Magnetics, turbulence, and cloud-cloud collisions probably have an effect on filamentary structure \citep{Jimenez2014,Smith2014,Planck2016a}. \citet{Jappsen2005} suggested that individual filaments can arise from compression of initially uniform gas by converging turbulent flows, while \citet{Vazquez2007} and \citet{Heitsch2008} argued that  filaments can radiate from hubs following the collision of uniform cylinders of gas.

On large scales, IRDCs are frequently associated with filamentary structures which are common features in giant molecular clouds (GMCs). Observations have revealed filaments with extents from a few to hundreds of parsecs \citep{Jackson2010,Li2013,Ragan2014,Goodman2014,Li2016,2016ApJS..226....9W}. These filaments are always with active star formation \citep{2010A&A...518L.102A,2014prpl.conf...27A,zhangcp2016}, and the large ones have been suggested as ``bones'' of the Milky Way \citep{Li2013,Li2016,2016ApJS..226....9W,Jackson2010,Li2013,Ragan2014,Goodman2014,Li2016,2016A&A...590A.131A,2015ApJ...815...23Z}. On small scales, IRDCs always fragment into dense clumps and cores which are the immediate birthplaces of stars. With low temperatures and high densities \citep{Egan1998,Carey1998,Pillai2006}, IRDCs may represent an early phase of star cluster formation \citep{Rathborne2006,Chambers2009,Sanhueza2012}. Recent studies have revealed massive (10 $\Msun$) young stellar objects (YSOs) \citep{Rathborne2005,Pillai2006,Beuther2007} and massive (10 - 1000 $\Msun$) cores in IRDCs \citep{Rathborne2006,Rathborne2007,Rathborne2011,Henning2010}. Mainly identified by their infrared or maser emission \citep{Dirienzo2015}, YSOs in IRDCs could be at very early stages of star formation. Some of the coldest and densest cores deeply embedded in IRDCs are just beginning the process of fragmentation and condensation.

The cold and dense features also lead to different chemistry in IRDCs than in other star-forming regions. For instance, CO, which is a frequently used tracer of dynamics in dense and cold clouds \citep{Hernandez2011}, could be affected by depleted gas phase abundances due to freezing out onto dust grains, especially in the coldest and densest regions \citep{Hernandez2011a}. By contrast, some N-bearing and D-bearing species can be tracers of choice for dense core gas due to their resistance to freeze out. Sometimes these species may also suffer from systematic abundance variations during core evolution due to enhancement by the disappearance of CO from the gas phase \citep{Caselli1999,Bergin2002,Tafalla2002,Aikawa2005,Bergin2007}. As suggested in the literature, these chemical features (e.g., depletion, deuteration) may indicate evolutionary stages in IRDCs \citep{Kong2015,Kong2016}.

In this paper, we have carried out observations of six sources in lines of C$^{18}$O (1-0), HCO$^+$ (1-0), HNC (1-0), N$_2$H$^+$ (1-0), DCO$^+$ (3-2), DCN (3-2), and SiO (5-4) using the IRAM 30-m and CSO 10.4-m telescopes. These species have different levels of depletion and varying critical densities. Combining with the archival data in the infrared and sub-millimeter, we can efficiently study dynamical and chemical properties of the targets of interest. This paper is arranged as follows: The observations and archival data in Section \ref{sect:data}, and the results in Section \ref{sect:results}. In Section \ref{sect:discu}, we discuss the dynamics, chemistry, and evolution of IRDCs. Finally, a summary is presented in Section \ref{sect:summary}.

\section{Observation and data processing}
\label{sect:data}

\subsection{Observational samples}
\label{sect:sources}

We selected (without bias) six samples in \citet{Rathborne2006}, and carried out multi-wavelength observations on IRDCs G031.97+00.07 \citep[G31;][]{Dirienzo2015}, G033.69-00.01 \citep[G33;][]{Sanhueza2012}, G034.43+00.24 \citep[G34;][]{Rathborne2005,Rathborne2008,Rathborne2011,Foster2014,Yanagida2014,Dirienzo2015,Pon2015,Pon2016,Xu2016}, G035.39-00.33 \citep[G35;][]{Henshaw2013,Henshaw2014,Jimenez2014,Dirienzo2015}, G038.95-00.47 \citep[G38;][]{Dirienzo2015}, and G053.11+00.05 \citep[G53;][]{Kim2015}. There have already been extensive studies  on the samples of IRDCs G34 and G35, while there are just few research works dedicated to the other IRDCs. For example, the IRDC G34 has been investigated with $\co$ (3-2), HCN (4-3), CS (3-2), and SiO (2-1) on a large scale by \citet{Rathborne2005}. However, a homogeneous and simultaneous analysis of HCO$^+$, HNC, N$_2$H$^+$, and C$^{18}$O for this sample is still missing. Combining the selected tracers HCO$^+$, HNC, N$_2$H$^+$, and C$^{18}$O, they can better probe dense and cold conditions in chemistry and dynamics than previous works.

\subsection{IRAM 30 m observations}
\label{sect:data_iram30}

Observations of the IRDCs G31, G33, G34, G35, G38, and G53, the HCO$^+$ (1-0), HNC (1-0), N$_2$H$^+$ (1-0), and C$^{18}$O (1-0) were carried out during December 2013 and April 2014 using the IRAM 30 m telescope\footnote{Based on observations carried out with the IRAM 30 m Telescope. IRAM is supported by INSU/CNRS (France), MPG (Germany) and IGN (Spain).} on Pico Veleta, Spain. The E90 band of the new Eight MIxer Receiver (EMIR) simultaneously covered the four lines. The Fast Fourier Transform Spectrometer (FFTS) backends were set at 50 kHz (about 0.15 $\kms$ at 90 GHz) resolution.

For the HCO$^+$, HNC, N$_2$H$^+$, and C$^{18}$O lines in the observations, the half-power beam width (HPBW) was between 29.0$''$ and 23.5$''$, the main beam efficiency ($B_{\rm eff}$) was between 81\% and 78\%, and the forward efficiency ($F_{\rm eff}$) of the IRAM 30 m telescope was between 95\% and 94\%, respectively. The relation between main beam temperature ($T_{\rm MB}$) and antenna temperature ($T^*_{\rm A}$) is $T_{\rm MB} = (F_{\rm eff} / B_{\rm eff}) \times T^*_{\rm A}$. The on-the-fly mapping mode was used to scan each sample in two orthogonal directions, to reduce striping on the maps. The sampling step was set as 9.3$''$, which meets the Nyquist sampling theorem for these species.

Calibration scans, pointing, and focus were done on a regular basis to assure correct calibration. Calibration scans were done at the beginning of each subscan. A pointing was done every hour, appriximately, and a focus scan every three hours, with more scans performed around sunset and sunrise. A clean reference position of each source was selected, and position switching was adopted to remove background noise. The flux calibration is expected to be accurate within 10\%. The GILDAS\footnote{http://www.iram.fr/IRAMFR/GILDAS/} and MIRIAD\footnote{http://www.cfa.harvard.edu/sma/miriad/} software packages were used to reduce the observational data.

The spectral profiles, especially for $\hcop$ and HNC, are strongly absorbed at the line center. If we assume that the spectral features are mostly from self-absorption, we can mask the absorption dip to make Gaussian fitting using CLASS package in GILDAS software, so that we can estimate their ``true'' line width and intensity. Additionally, for  core G31-8, for example, the blueshifted wing of $\co$ was not well fitted, so its Gaussian fitting line-center velocity is not consistent with the spectra $\hcop$, HNC, and $\nhp$. The Gaussian fitting images of the lines HCO$^+$, HNC, N$_2$H$^+$, and C$^{18}$O are shown in Fig. \ref{Fig:iram-spectra}, and the fitted parameters are listed in Table \ref{tab:hcop}.

\subsection{Observations of CSO 10.4 m}

We also report the deep single-point detection of the lines DCO$^+$ (3-2) at 216.113 GHz, SiO (5-4) at 217.105 GHz, DCN (3-2) at 217.239 GHz in the six IRDCs using Caltech Submillimeter Observatory (CSO) 10.4 m telescope\footnote{This material is based upon work at the Caltech Submillimeter Observatory, which is operated by the California Institute of Technology.}. The observations singly point to the peak positions of the densest cores extracted from the IRDCs (see Section \ref{sect:seletion}). A clean reference of each source makes sure that position switching mode was used to remove background. One 230 GHz receiver at the CSO is a double-side band (DSB) receiver. These lines were simultaneously covered by Fast Fourier Transform Spectrometer (FFTS). This FFTS backend provides two 4-GHz windows of DSB coverage with a frequency resolution of 269 kHz. The data processing was made with the GILDAS/CLASS software. The full width at half medium (FWHM) beam size on Mars was about 33.4$''$ at 230 GHz. The data was converted to $T_{\rm MB}$ using a beam efficiency of 69\%, which was measured toward Jupiter. The Gaussian fitting parameters of the detected four lines are listed in Table \ref{tab_cso-spectra}.

\subsection{Archival data}
\label{sect:archive}

The combined data comprise \textit{Spitzer} IRAC 8.0 $\mu$m \citep[$1\sigma$ = 78~$\mu$Jy;][]{benj2003,chur2009}, MIPSGAL 24 $\mu$m \citep[$1\sigma$ = 70~$\mu$Jy;][]{care2009}, PACS 70 and 160 $\mu$m \citep[$1\sigma_{70~{\mu}m}$ = 1~mJy, $1\sigma_{160~{\mu}m}$ = 2~mJy;][]{Poglitsch2010}, SPIRE 250, 350, and 500 $\mu$m \citep[$1\sigma_{250~{\mu}m}$ = 1.16~mJy, $1\sigma_{350~{\mu}m}$ = 1.26~mJy, and $1\sigma_{500~{\mu}m}$ = 1.36~mJy;][]{Griffin2010}, and ATLASGAL\footnote{The ATLASGAL project is a collaboration between the Max-Planck-Gesellschaft, the European Southern Observatory (ESO) and the Universidad de Chile.} 870 $\mu$m \citep[$1\sigma$ = 45 $\mjyb$;][]{Schuller2009}. The \textit{Herschel} Space Observatory is a 3.5 meter telescope observing the far-infrared and submillimeter universe. The InfraRed Array Camera (IRAC) is one of three focal plane instruments on the \textit{Spitzer} Space Telescope. The IRAC is a four-channel camera that provides simultaneous $5.2' \times 5.2'$ images at 3.6, 4.5, 5.8, and 8.0 $\mu$m. The Multiband Imaging Photometer for \textit{Spitzer} (MIPS) produced imaging and photometry in three broad spectral bands, centered nominally at 24, 70, and 160 $\mu$m, and low-resolution spectroscopy between 55 and 95 $\mu$m. The imaging bands for the Photo detector Array Camera and Spectrometer (PACS) were centered at 70, 100, and 160 $\mu$m. ATLASGAL is the APEX Telescope Large Area Survey of the Galaxy, an observing program with the LABOCA bolometer array at APEX, located at 5100 m altitude on Chajnantor, Chile. Its spatial resolution at 870 $\mu$m is about 19$''$.

\section{Results and analysis}
\label{sect:results}

\subsection{Core selection}
\label{sect:seletion}

A typical terminology \citep[e.g.,][]{Bergin2007} for cloud, clumps, and cores has a physical size of 2 -- 15, 0.3 -- 3, and 0.03 -- 0.2 pc, and mass of about 10$^3$ -- 10$^4$, 50 -- 500, and 0.5 -- 5 $\Msun$, respectively. However, in this work the six IRDCs are selected from \citet{Rathborne2006}, who showed that the extracted objects are cores, although some of them (shown in Table \ref{tab:n2hp}) are in clump size. To be consistent, we follow the definition in \citet{Rathborne2006} to investigate the identified cores within the six IRDCs.

\citet{Rathborne2006} identified 47 cores with 1.2 mm continuum data using IRAM 30m telescope toward the six IRDCs. Their FWHM angular resolution is 11$''$. In addition, ATLAGAL 870 $\mu$m continuum has a relatively low spatial resolution of 19$''$. To contrast with the work of \citet{Rathborne2006}, we will use their extracted core parameters (including core positions and core sizes) for further comparative investigation.

Among the 47 cores identified, ten cores are not covered by our HCO$^+$, HNC, N$_2$H$^+$, and C$^{18}$O observations; in our samples, seven cores are not covered or distinguished by \citet{Rathborne2006}. We carefully extracted the seven cores by comparing 870 $\mu$m continuum and N$_2$H$^+$ integrated intensity emission. Finally, in total we obtained 44 cores shown in Table \ref{tab:n2hp}. These core numbers with boldface are newly identified cores in this work.

In the N$_2$H$^+$ and $\co$ maps of Fig. \ref{Fig:int-map}, these numbers indicate the selected core positions. These spectra (including HCO$^+$, HNC, N$_2$H$^+$, and C$^{18}$O) shown in Fig. \ref{Fig:iram-spectra} were extracted from the selected core positions within their own core sizes. In addition, the derived integrated fluxes of 870 $\mu$m and masses listed in Table \ref{tab:n2hp} were also estimated within the core sizes.

\subsection{Optical depth and line width}

Spectra N$_2$H$^+$ has seven hyperfine structure (HFS) lines. In the three distinguished emission lines of Fig. \ref{Fig:iram-spectra}, however, only the transition $1_{0,\,1} \rightarrow 0_{1,\,2}$ does not blend any other hyperfine line. Therefore, we can extract line width by setting this isolated transition component as systemic velocity when fitting HFS lines. The relative velocity and intrinsic intensity of the seven hyperfine lines are respectively (0.0 $\kms$, 1.00000), (7.394 $\kms$, 1.00000), (8.006 $\kms$, 2.33333), (8.961 $\kms$, 1.66667), (13.553 $\kms$, 1.00000), (13.987 $\kms$, 1.66667), and (14.942 $\kms$, 0.33334) \citep{Caselli1995,Keto2010}. Based on the parameters above, we use CLASS package in GILDAS software to fit these HFS lines. The green curves on spectra N$_2$H$^+$ in Fig. \ref{Fig:iram-spectra} show the fitting results, and the fitting parameters (such as line width $\Delta V_{\nhp}$ and optical depth $\tau_{\nhp}$), are listed in Table \ref{tab:n2hp}. We note that the cores G31-3, G35-4, and G35-5 were fitted using two velocity components. The derived optical depth mainly ranges from 0.10 to 0.24, so the N$_2$H$^+$ is optically thin in the condition of these dense and cold cores. Checking with RADEX online\footnote{http://var.sron.nl/radex/radex.php} \citep{Tak2007}, the line is indeed optically thin on the condition, which is basically consistent with other dense cores in IRDCs \citep[e.g.,][]{Caselli2002,Sanhueza2012,Liu2013}.

\subsection{PV diagram}

The molecular ion HCO$^+$ is often used as dynamical tracer of dense gas. In Fig. \ref{Fig:pv}, we show the position-velocity (PV) diagrams for IRDCs G31, G33, G34, G35, G38, and G53. The broken lines in the HCO$^+$ subfigure of Fig. \ref{Fig:int-map} show the cutting direction, which goes through the selected dense core positions. The dense cores are the backbone of IRDCs, so we can study the dynamical status of the IRDCs and their embedded cores. Based the PV diagrams in Fig. \ref{Fig:pv} and combining with optically thin line $\nhp$ in Fig. \ref{Fig:iram-spectra}, we can see that the IRDCs show features with multi-velocity component (e.g., G31-3), multi-core structure (e.g., G38 and G31), outflow (e.g., G34 and G53), and velocity gradient (e.g., G35).

\subsection{Dust temperature and column density}
\label{sect:temperature}

The high-quality Hi-GAL data cover a large wavelength range from 70 to 500 $\mu$m making it practical to obtain dust temperature maps of these IRDCs via fitting the spectral energy distribution (SED) to the multi-wavelength images on a pixel-by-pixel basis. Firstly we have followed \citet{Wang2015} to perform Fourier transfer (FT) based on background removal. In this method, the original images are firstly transformed into Fourier domain and separated into the low and high spatial frequency components, and then inversely Fourier transferred back into image domain. The low-frequency component corresponds to large-scale background and foreground emission, while the high-frequency component reserves the emission of interest. Detailed descriptions of the FT-based background removal method can be found in \citet{Wang2015}. After removing the background and foreground emission, we re-gridded the pixels onto the same scale of 11.5$''$, and convolved the images to a circular Gaussian beam with $\mathrm{FWHM = 45''}$ which corresponds to the measured beam of Hi-GAL observations at 500 $\mu$m \citep{Traficante2011}.

The intensities at multi-wavelengths of each pixel have been modeled as
        \begin{eqnarray}
        \begin{aligned}
        \label{equa_Su}
S_\nu=B_\nu(T) (1-e^{-\tau_\nu})
        \end{aligned}
        ,\end{eqnarray}
where the Planck function $B_\nu(T)$ is modified by optical depth
        \begin{eqnarray}
        \begin{aligned}
        \label{equa_tau}
\tau_\nu = \mu_\mathrm{H_2}m_\mathrm{H}\kappa_\nu N_\mathrm{H_2}/R_\mathrm{gd}.
        \end{aligned}
        \end{eqnarray}
Here, $\mu_\mathrm{H_2}=2.8$ is the mean molecular weight adopted from \citet{Kauffmann2008}, $m_\mathrm{H}$ is the mass of a hydrogen atom, $N_\mathrm{H_2}$ is the column density, $R_\mathrm{gd}=100$ is the gas to dust ratio. The dust opacity $\kappa_\nu$ can be expressed as a power law of frequency with
        \begin{eqnarray}
        \begin{aligned}
        \label{equa_kappa}
\kappa_\nu=5.0\left(\frac{\nu}{600~\mathrm{GHz}}\right)^\beta~\mathrm{cm^2g^{-1}},
        \end{aligned}
        \end{eqnarray}
where $\kappa_\nu(\mathrm{600~GHz})=5.0~\mathrm{cm^2g^{-1}}$ adopted from \citet{Ossenkopf1994}. The dust emissivity index has been fixed to be $\beta=1.75$ according to \citet{Battersby2011}. The free parameters are the dust temperature $T_{\rm dust}$ and column density $N_\mathrm{H_2}$.

The final resulting dust temperature and column density maps, which have a spatial resolution of 45$''$ with a pixel size of 11.5$''$, are shown in Figs. \ref{Fig:temperature} and \ref{Fig:column_density}, respectively. The dust temperatures range from 14.03 K to 24.37 K (see Table \ref{tab:n2hp}) with minimums associated with N$_2$H$^+$ peaks, while the derived column densities range from about $1.0 \times 10^{22}$ to $1.4 \times 10^{23}$ cm$^{-2}$, where the maximum is located at the peak of core G34-1. The estimated dust temperatures and column densities suggest that most of the cores embedded in IRDCs are cold and dense precursors to very early high-mass star formation.

\citet{Nguyen2011} have presented the temperature distribution of G35 derived from SED fitting combining the PACS and SPIRE data. The temperature distribution is similar to that in Fig. \ref{Fig:temperature}, which indeed indicates a cold filament. Roughly comparing the dust temperature in this work with that in \citet{Rathborne2005}, for example in G34, we find the dust temperatures $T_{\rm dust}$ for the cores G34-1, G34-3, and G34-4 are 19.7, 16.3, and 17.6 K, respectively. In \citet{Rathborne2005}, the corresponding cores have $T_{\rm dust}$ = 34, 32, and 32 K. The derived dust temperatures for G34 in this work  have a factor of about two times lower than those in \citet{Rathborne2005}. The cores may be constituted by a small-scale hot component and a large-scale thick cold envelope \citep{Rathborne2005,Rathborne2011}. The different photometry aperture and spatial resolution may lead to different dust temperatures in the two works.

\subsection{Mass and volume density}
\label{sect:mass}

In this section, we use 870 $\mu$m data only, along with the derived dust temperatures in Section \ref{sect:temperature} to estimate masses and volume densities of these dense cores, considering its higher spatial resolution and better capability than \textit{Herschel} data in tracing cold cores. It is assumed that the dust emission is optically thin and the gas to dust ratio is 100. The core masses are calculated using dust opacities of 0.0185 cm$^2$ g$^{-1}$ at 870 $\mu$m \citep{Ossenkopf1994}. The total masses $M_{\rm 870\,{\mu}m}$ of the sources can therefore be calculated
\citep{Kauffmann2008} through
        \begin{eqnarray}
        \begin{aligned}
        \label{equa_mass}
        \left(\frac{M_{\rm 870\,{\mu}m}}{\Msun}\right)   =  & 0.12 \left({\rm  e}^{14.39\left(\frac{\lambda}{\rm mm}\right)^{-1}\left(\frac{T_{\rm dust}}{\rm K}\right)^{-1}}-1\right) \times &  \\
        & \left(\frac{\kappa_{\nu}}{\rm cm^{2}g^{-1}}\right)^{-1} \left(\frac{S_{\nu}}{\rm Jy}\right) \left(\frac{D}{\rm kpc}\right)^{2}
        \left(\frac{\lambda}{\rm mm}\right)^{3}, &
        \end{aligned}
        \end{eqnarray}
where $\lambda$ is the observational wavelength, $T_{\rm dust}$ is the dust temperature, $\kappa_{\nu}$ is the dust opacity, $S_{\nu}$ is the integrated flux, and $D$ is the distance to the Sun. If the cores are considered as an uniform sphere, the volume density $n_{\rm 870\,{\mu}m}$ can be roughly estimated by $n_{\rm 870\,{\mu}m} = M_{\rm 870\,{\mu}m}/{(2R_{\rm core})}^3$. These corresponding and derived parameters are listed in Table \ref{tab:n2hp}.

The virial theorem can be used to test whether a molecular fragment is in a stable state. Under the assumption of a simple spherical cloud with a density distribution of $\rho = constant$, if we ignore magnetic fields and bulk motions of the gas \citep{MacLaren1988,Evans1999},
        \begin{eqnarray}
        \begin{aligned}
        \label{equa_virial-mass}
    M_{\rm vir} \simeq 210\, r\, \Delta V^{2}\, (\Msun),
        \end{aligned}
        \end{eqnarray}
where $r$ is the fragment radius in pc and $\Delta V$ is the FWHM line width in $\kms$. The virial parameter $\alpha_{\rm vir}$ is defined by $\alpha_{\rm vir} = M_{\rm vir}/M_{\rm 870\,{\mu}m}$. These parameters are also listed in Table \ref{tab:n2hp}.

Magnetic field might play a crucial role in IRDC dynamics \citep{Li2014,Pillai2015}. When $\alpha_{\rm vir} = M_{\rm vir}/M_{\rm 870\,{\mu}m} \ll 1$, one would expect additional factors to work against collapse, and magnetic field would be a good candidate. Among our sample, for most of the cores in IRDCs G31 and G34, the virial parameters are significantly smaller than one. This would suggest that magnetic field might play a crucial role here. These sources would be ideal targets for further studies.

\subsection{Diagrams of integrated intensity and column density}

\begin{figure*}
\centering
\includegraphics[width=0.35\textwidth, angle=0]{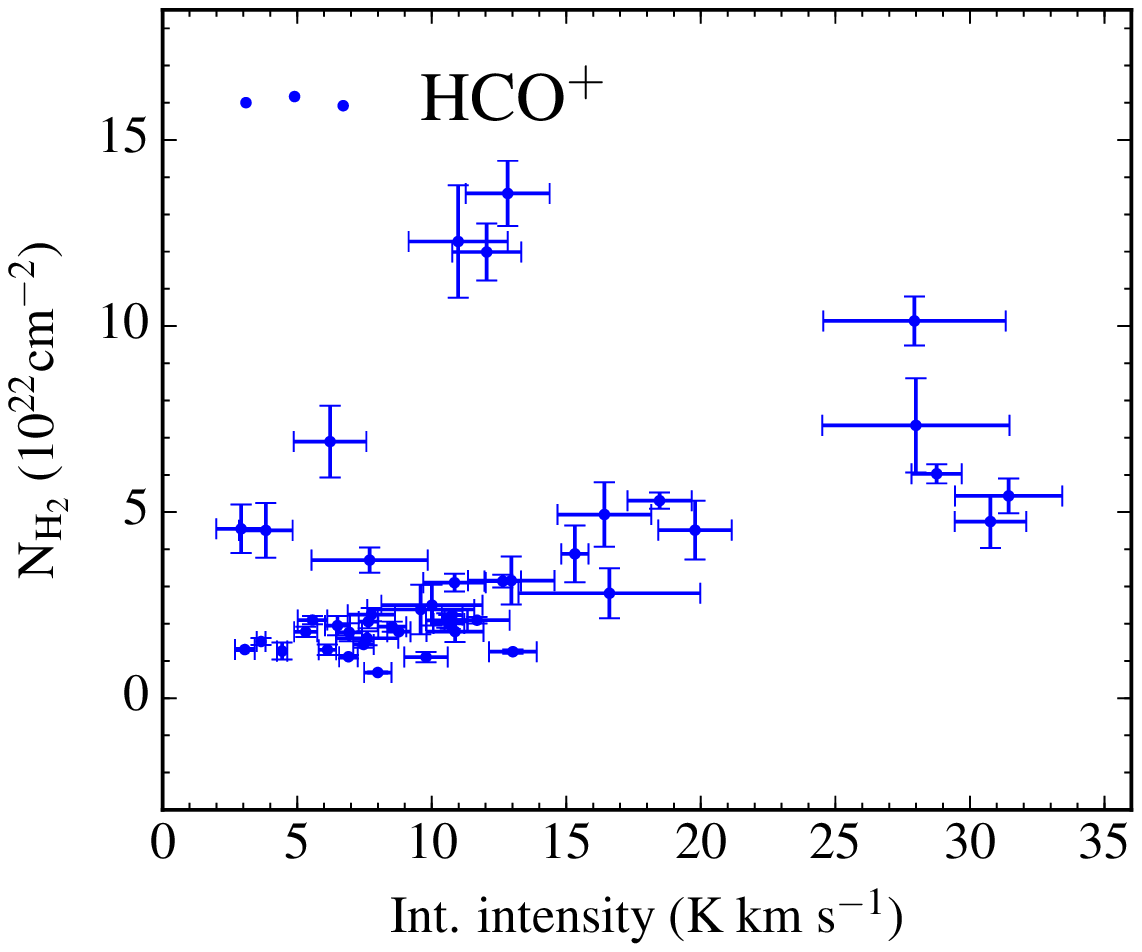}
\includegraphics[width=0.35\textwidth, angle=0]{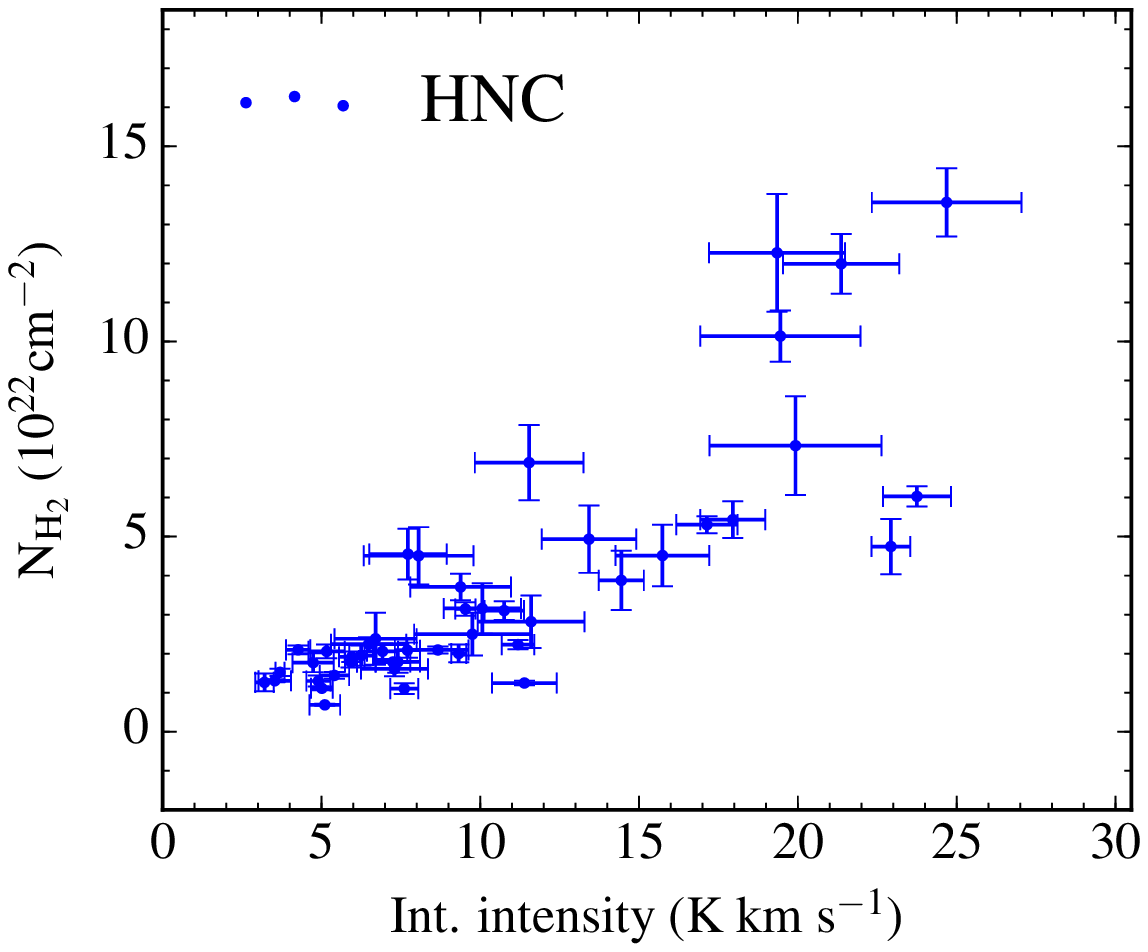}
\includegraphics[width=0.35\textwidth, angle=0]{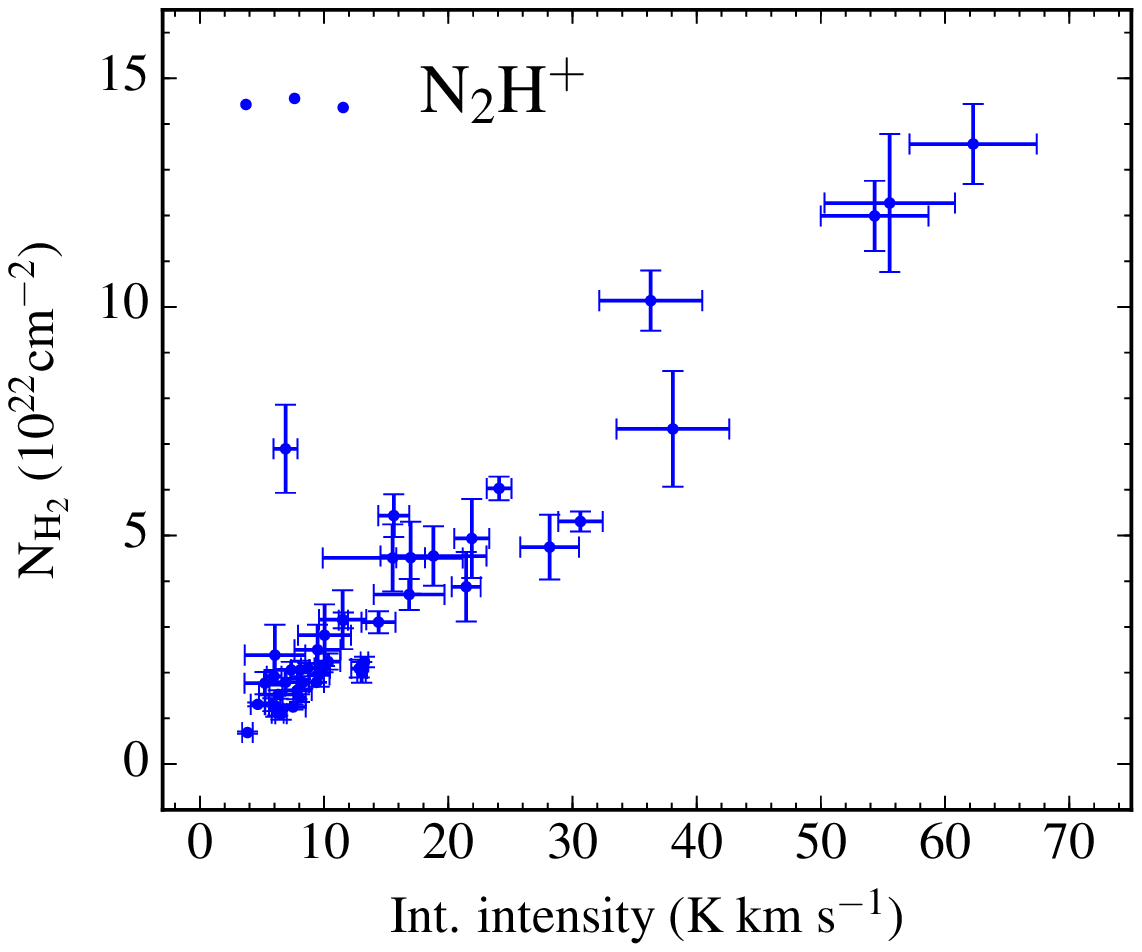}
\includegraphics[width=0.35\textwidth, angle=0]{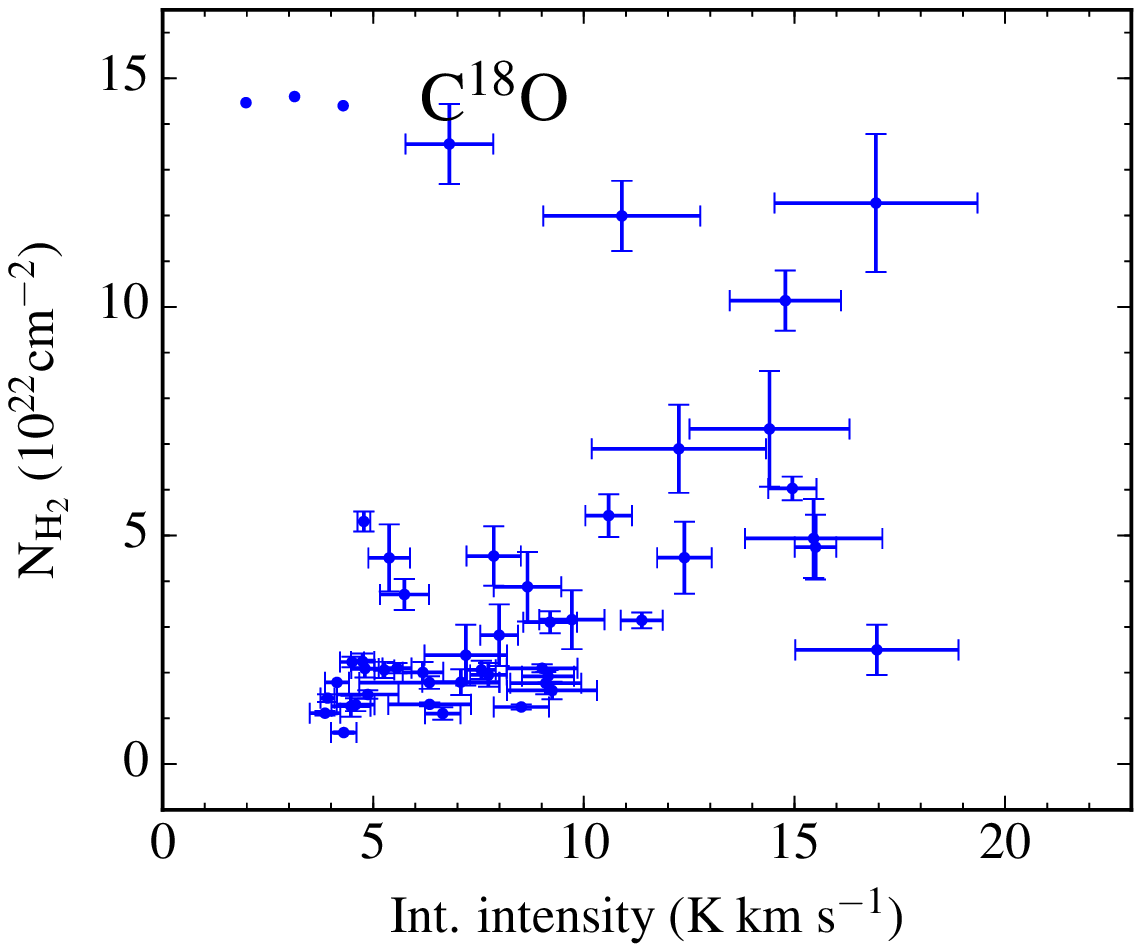}
\caption{Column density and integrated intensity diagrams between  $\hcop$, HNC, $\nhp$, and $\co$. Column density was derived from SED fits with \textit{Herschel} data (see Section \ref{sect:temperature}). The data points are derived from the average within one beam size ($\sim27''$) with rms > 6$\sigma$, and selected without bias along the six IRDCs.}
\label{Fig:N-int}
\end{figure*}

\begin{figure*}
\centering
\includegraphics[width=0.33\textwidth, angle=0]{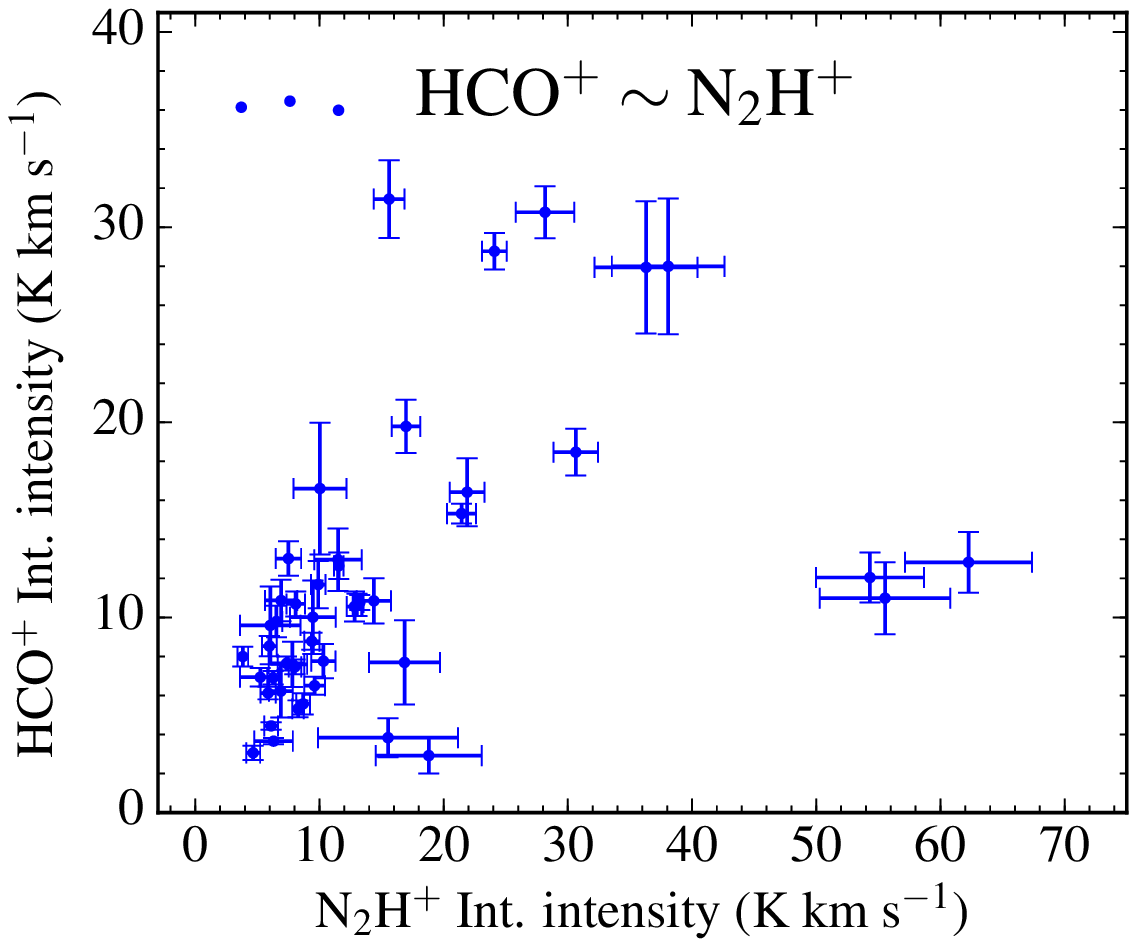}
\includegraphics[width=0.33\textwidth, angle=0]{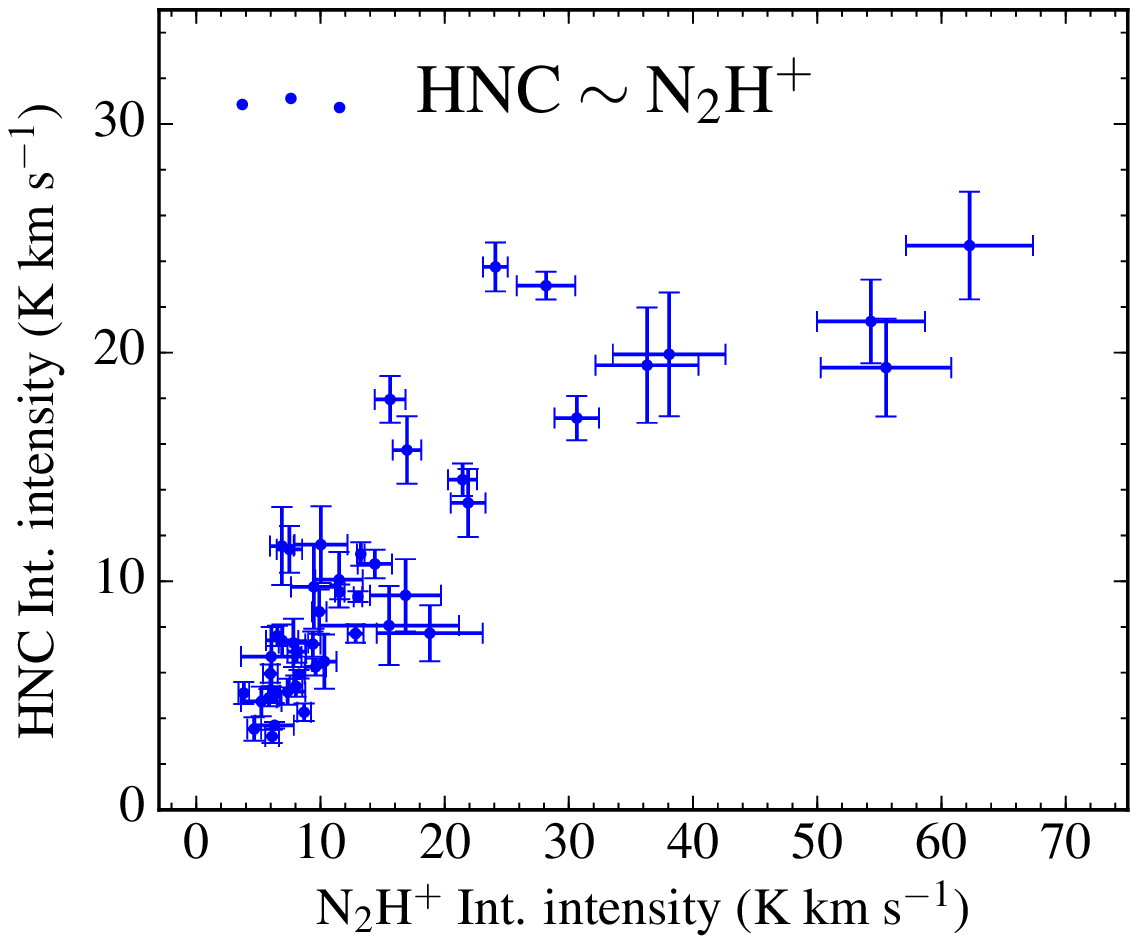}
\includegraphics[width=0.33\textwidth, angle=0]{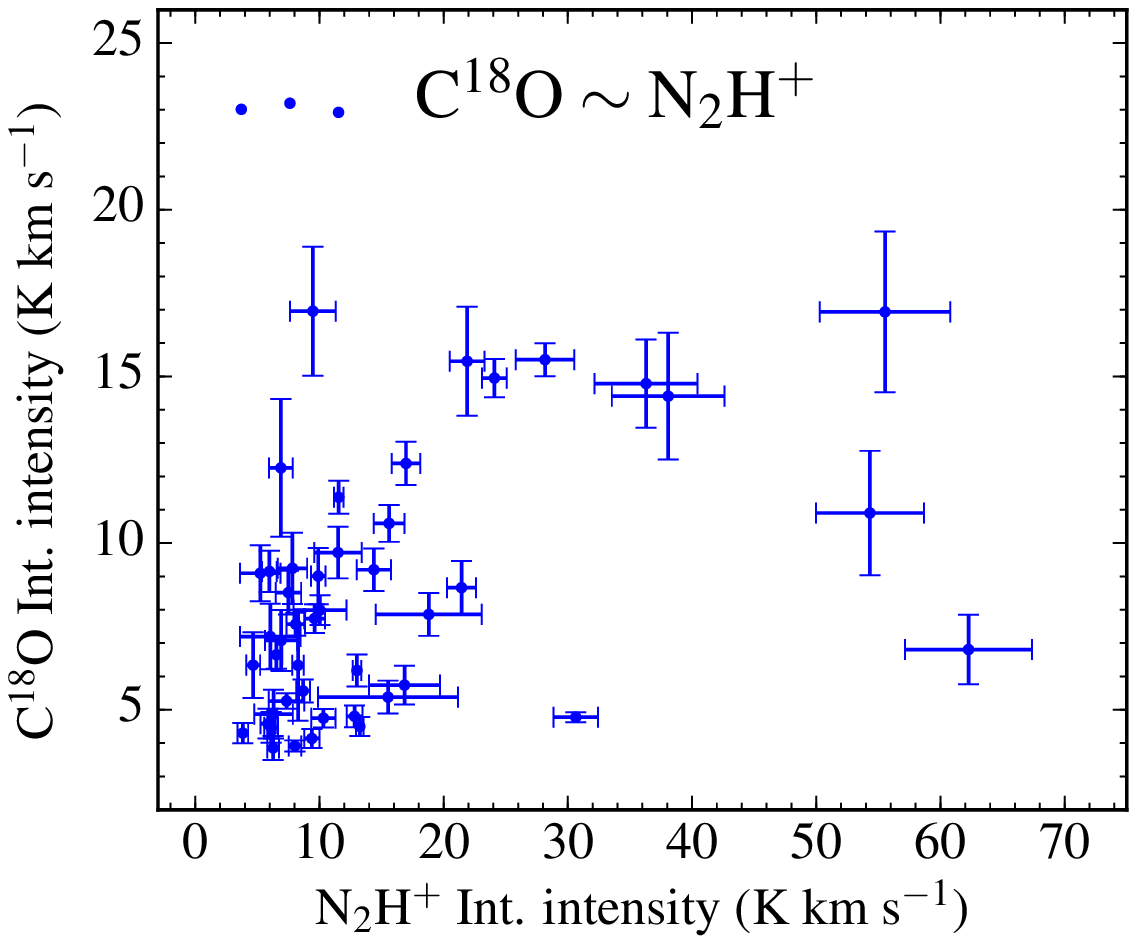}
\caption{Integrated intensity diagrams between $\hcop$, HNC, $\nhp$, and $\co$. The data points are derived from the average within one beam size ($\sim27''$) with rms > 6$\sigma$, and selected without bias along the six IRDCs.}
\label{Fig:int-int}
\end{figure*}

Figures \ref{Fig:N-int} and \ref{Fig:int-int} show diagrams of the relations between integrated intensity and column density. The integrated intensities of $\hcop$, HNC, $\nhp$, and $\co$ are derived from the average within one beam size ($\sim27''$) with rms > 6$\sigma$, and are selected without bias along the six IRDCs. To probe the spatial distribution of species ($\hcop$, HNC, $\nhp$, and $\co$) and column density, we use only the integrated intensities of the species instead of converting them into column density, which would introduce some uncertain assumption. In addition, the spatial scales between integrated intensity and column density are a little different, due to a different spatial resolution of data. This is acceptable given that we are mainly interested in statistical relationships.

\section{Discussion}
\label{sect:discu}

\subsection{The morphology of the IRDCs}

\begin{figure*}
\centering
\includegraphics[width = 0.99 \textwidth]{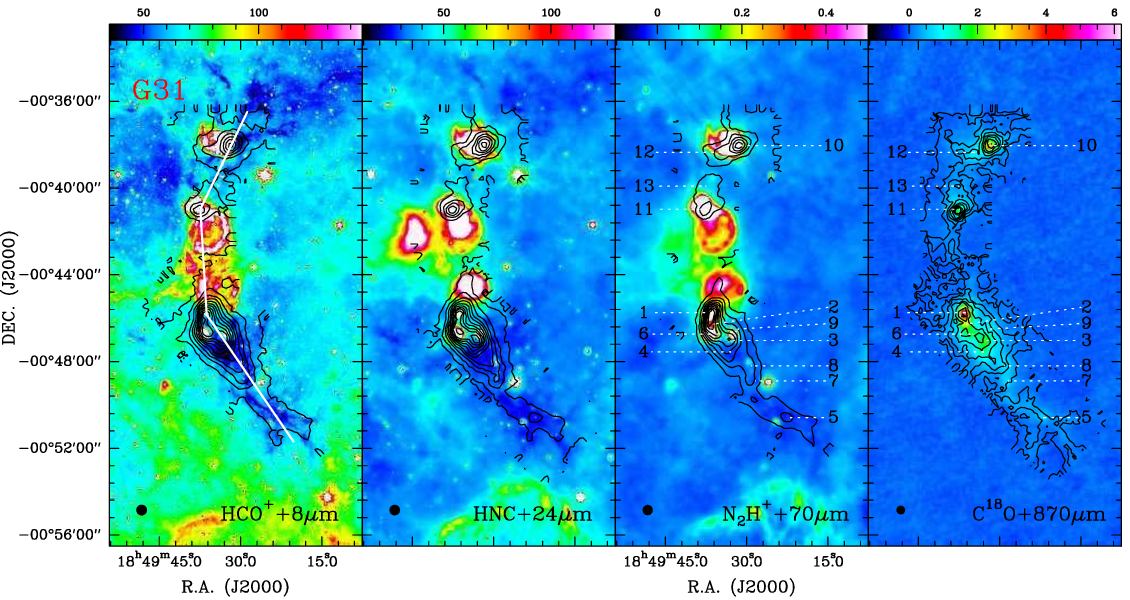}
\includegraphics[width = 0.99 \textwidth]{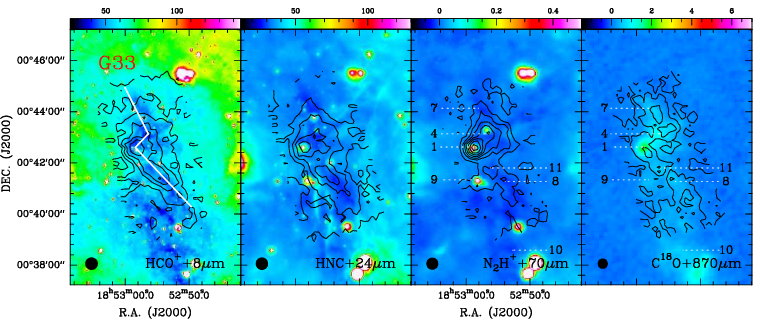}
\caption{Integrated intensity maps of the HCO$^+$, HNC, N$_2$H$^+$, and C$^{18}$O lines superimposed on 8, 24, 70, and 870 $\mu$m emission toward IRDCs G31, G33, G34, G35, G38, and G53. Their integrated velocity ranges are (80, 110), (92, 117), (46, 70), (31, 52), (35, 57), and (12, 34) $\kms$, respectively. For HCO$^+$, HNC, N$_2$H$^+$, and C$^{18}$O in these IRDCs, the contour levels start at 3 -- 5$\sigma$ in steps of 2 -- 6$\sigma$, with $\sigma$ = 0.75 -- 1.10 ${\rm K}\, \kms$. The beam size of each subfigure is indicated at the bottom-left corner. The numbers indicate the positions of the selected cores, which are the same as those in \citet{Rathborne2006}. The straight lines in the HCO$^+$ subfigure show the cutting direction of the PV diagram in Fig. \ref{Fig:pv}. The unit of each color bar is in MJy sr$^{-1}$ for 8, 24, and 70 $\mu$m, and in $\jyb$ for the 870 $\mu$m, respectively.}
\label{Fig:int-map}
\end{figure*}

\begin{figure*}
\centering
\includegraphics[width=0.95\textwidth, angle=0]{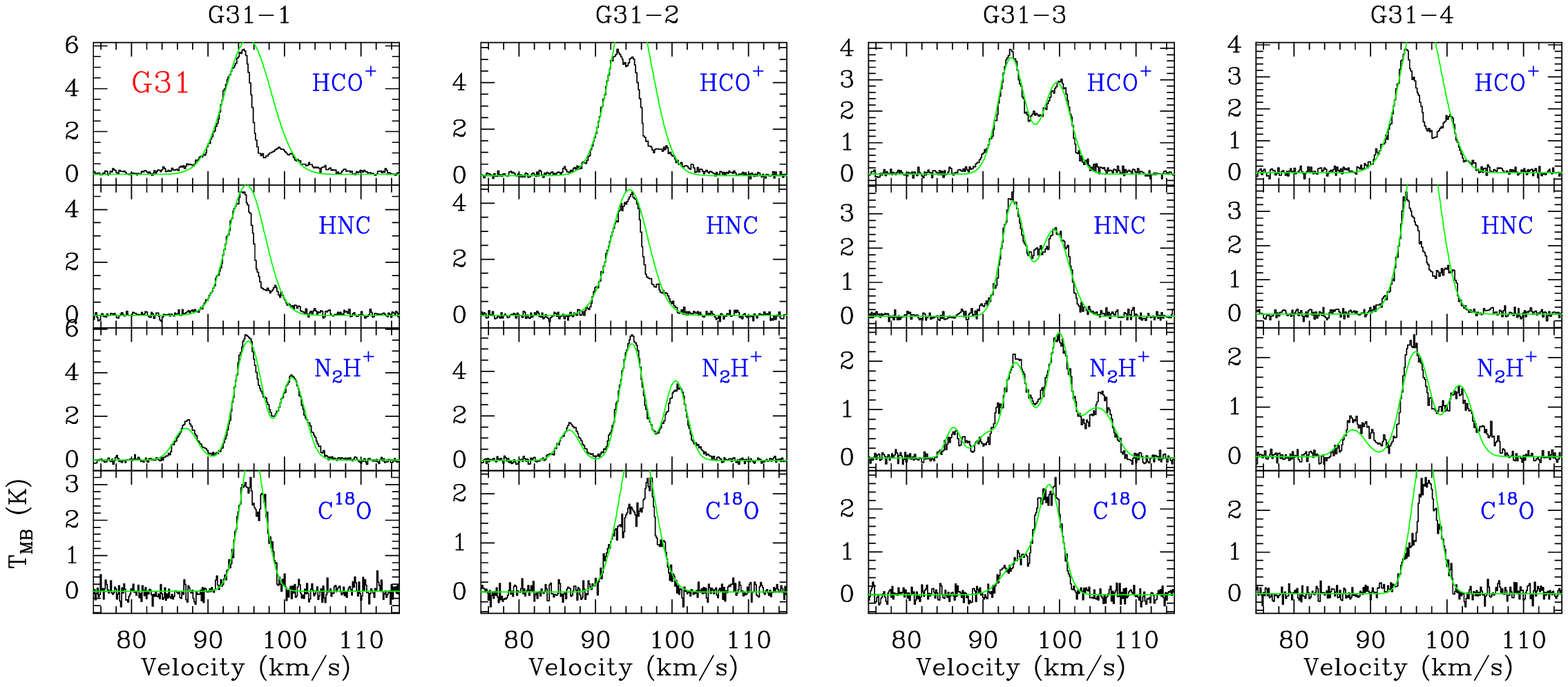}
\includegraphics[width=0.95\textwidth, angle=0]{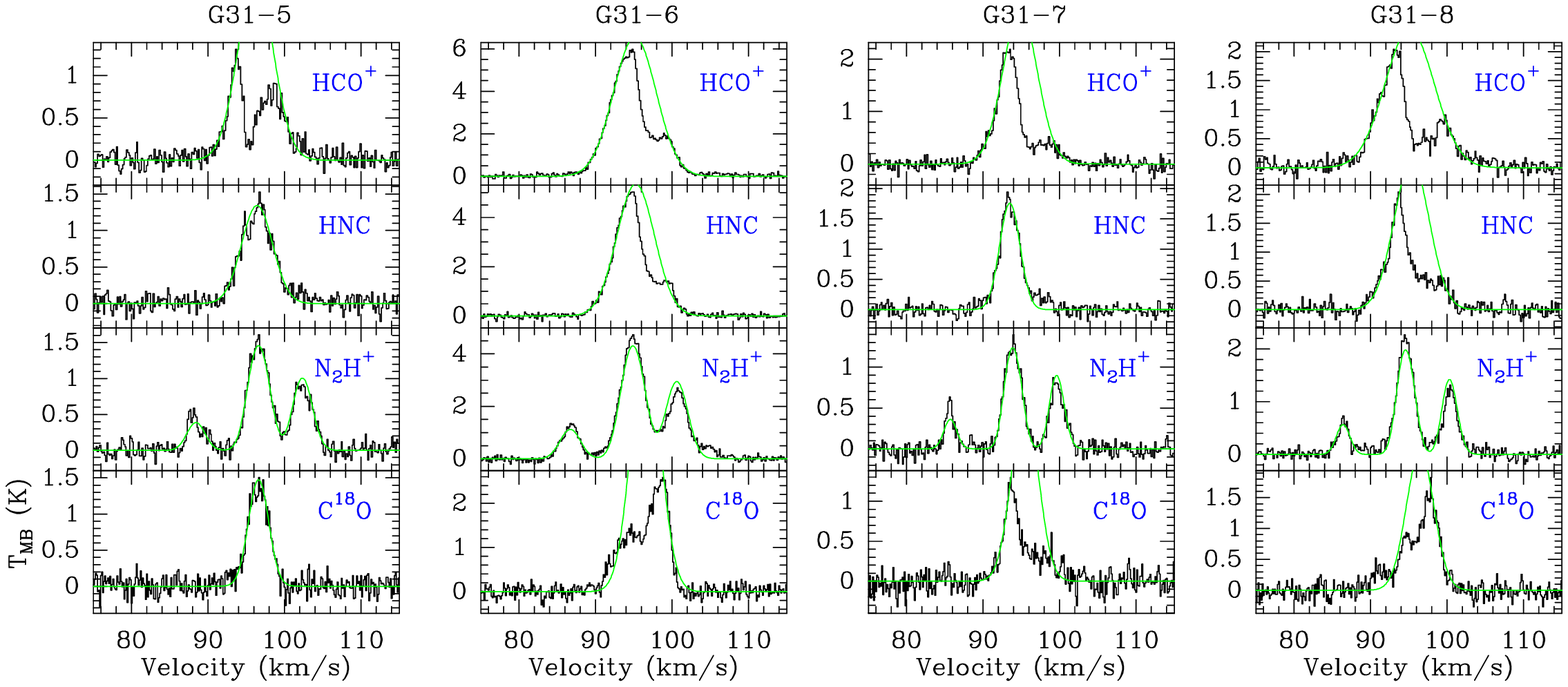}
\includegraphics[width=0.95\textwidth, angle=0]{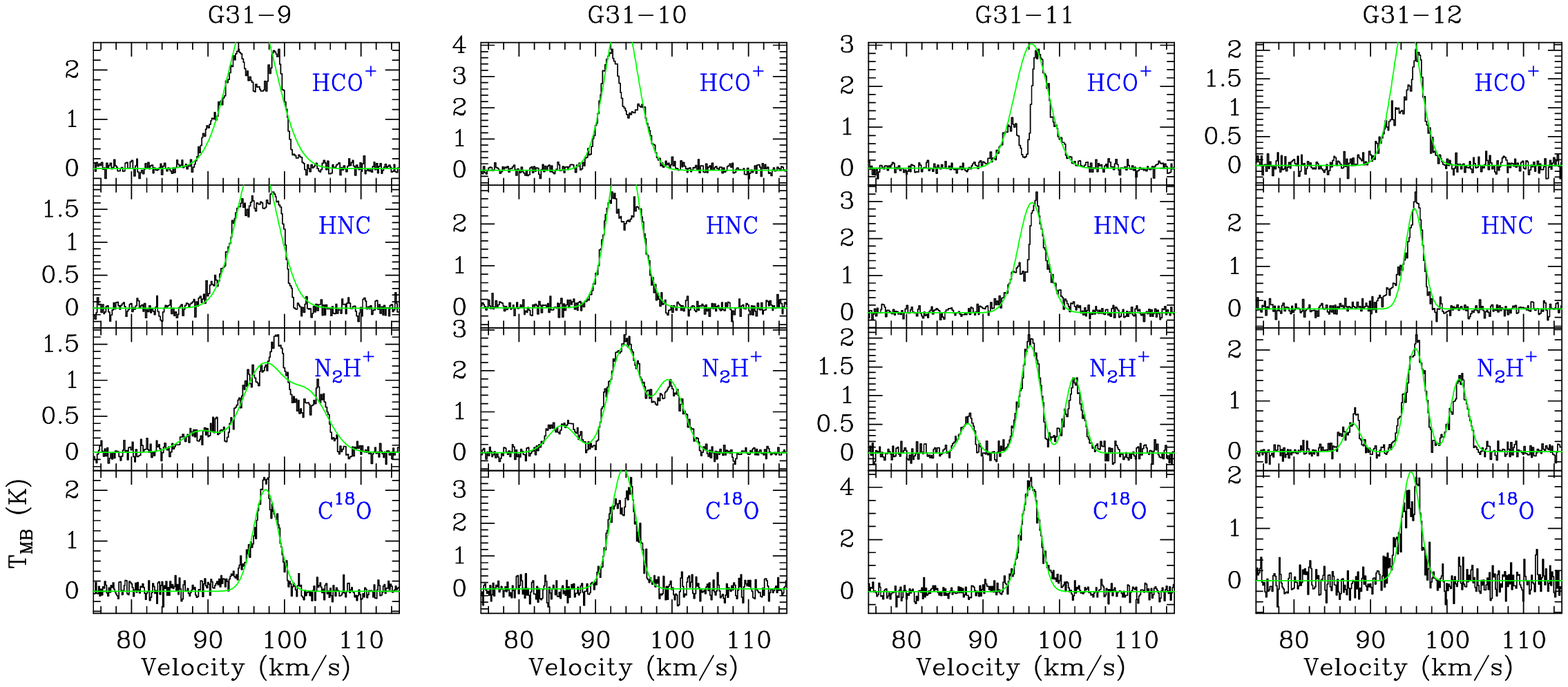}
\caption{Spectra of HCO$^+$, HNC, N$_2$H$^+$, and C$^{18}$O extracted from each core in Fig. \ref{Fig:int-map}. These spectra are the average within one beam size ($\sim 27''$). The green curves on N$_2$H$^+$ are from the HFS fit, and on other spectra are from the Gaussian fit by masking the absorption dip with GILDAS/CLASS software. To better show the HFS fit, the strongest HFS line was set the systemic velocity component.}
\label{Fig:iram-spectra}
\end{figure*}

\begin{figure*}
\centering
\includegraphics[width=0.30\textwidth, angle=0]{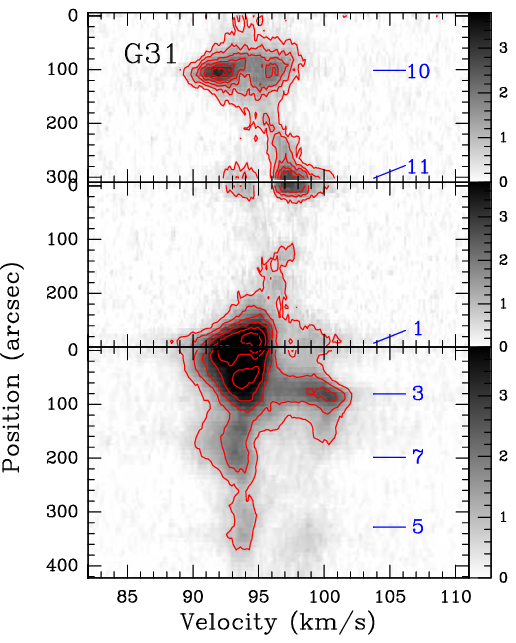}
\includegraphics[width=0.30\textwidth, angle=0]{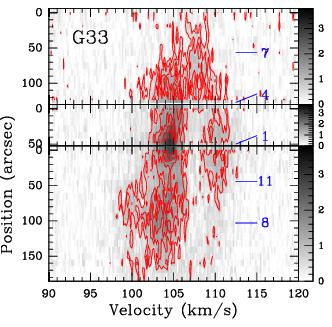}
\includegraphics[width=0.30\textwidth, angle=0]{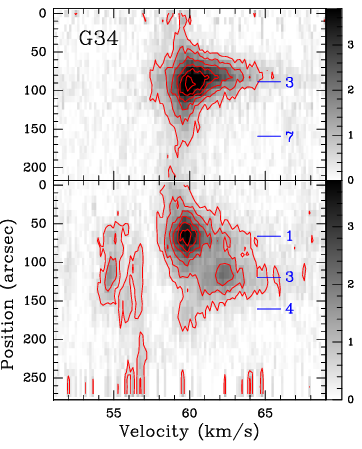}
\includegraphics[width=0.30\textwidth, angle=0]{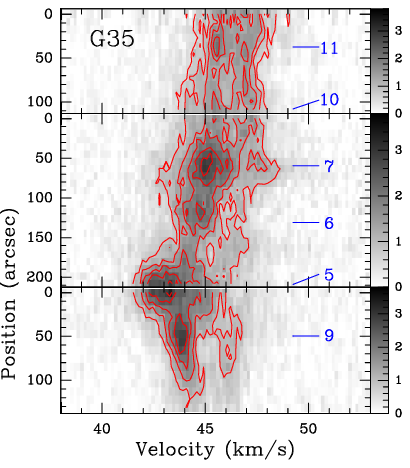}
\includegraphics[width=0.30\textwidth, angle=0]{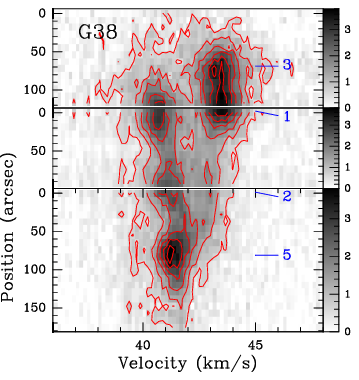}
\includegraphics[width=0.30\textwidth, angle=0]{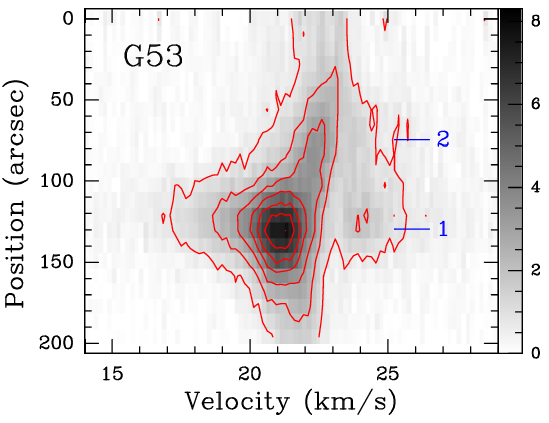}
\caption{Position-velocity diagrams with molecular ion $\hcop$ along the straight line (from the upper to lower) of each IRDC in Fig. \ref{Fig:int-map}. The numbers indicate the core positions corresponding to those in Fig. \ref{Fig:int-map}. The outer contour levels are above $3\sigma$ ($\sigma \rm \sim 0.16 \,K$). The unit of each color bar
is in K.}
\label{Fig:pv}
\end{figure*}

\begin{figure*}
\centering
\includegraphics[width=0.33\textwidth, angle=0]{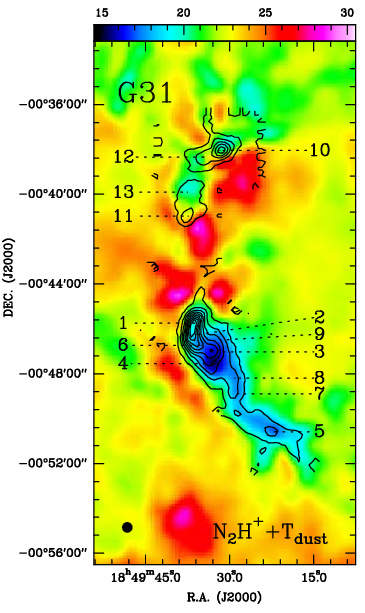}
\includegraphics[width=0.33\textwidth, angle=0]{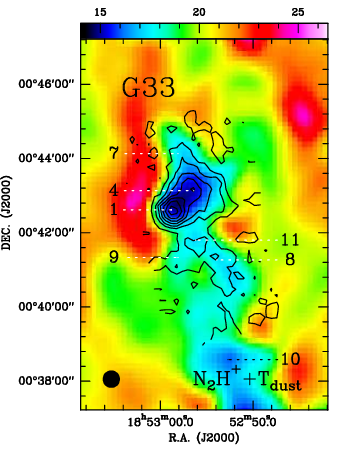}
\includegraphics[width=0.33\textwidth, angle=0]{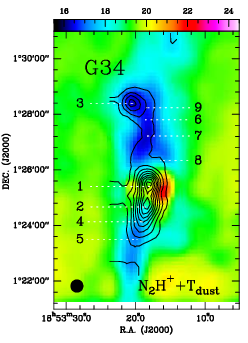}
\includegraphics[width=0.26\textwidth, angle=0]{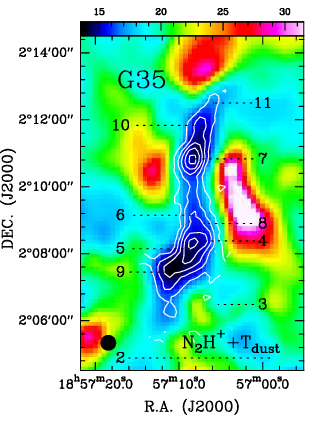}
\includegraphics[width=0.365\textwidth, angle=0]{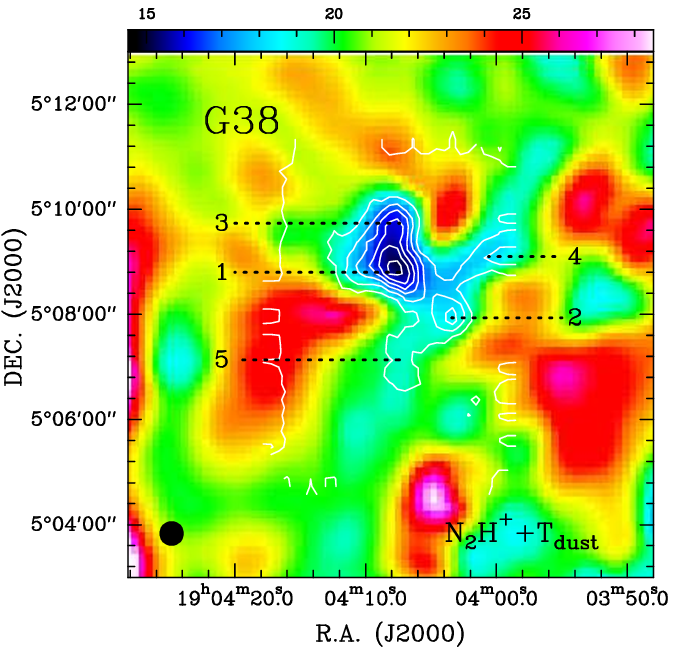}
\includegraphics[width=0.365\textwidth, angle=0]{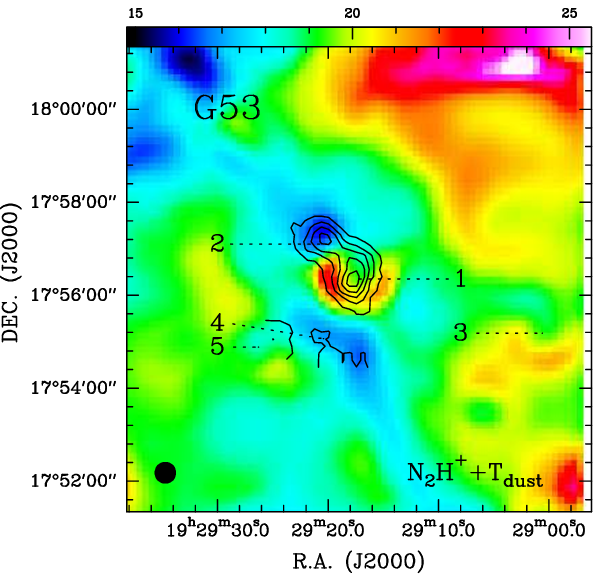}
\caption{The distributions of dust temperature ($T_{\rm dust}$) in color scale derived from SED fits pixel by pixel combining with \textit{Herschel} 70, 160, 250, 350, and 500 $\mu$m. The contours show the corresponding integrated intensities of N$_2$H$^+$ emission. The numbers indicate the positions of the selected cores, which are the same as those in \citet{Rathborne2006}. Contour levels and beam sizes are the same as those in Fig. \ref{Fig:int-map}. The unit of each color bar is in K.}
\label{Fig:temperature}
\end{figure*}

\begin{figure*}
\centering
\includegraphics[width=0.33\textwidth, angle=0]{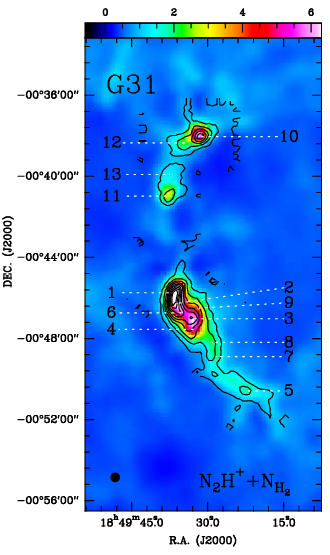}
\includegraphics[width=0.33\textwidth, angle=0]{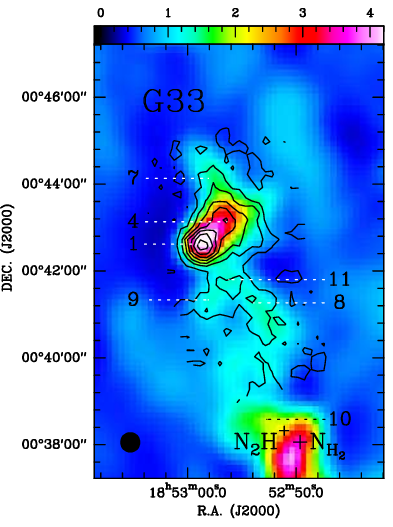}
\includegraphics[width=0.33\textwidth, angle=0]{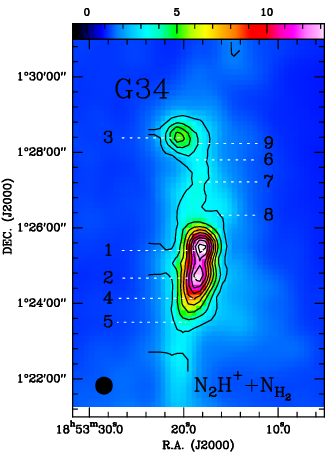}
\includegraphics[width=0.26\textwidth, angle=0]{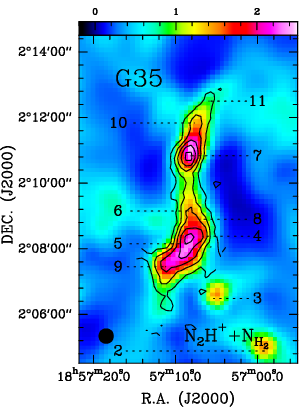}
\includegraphics[width=0.365\textwidth, angle=0]{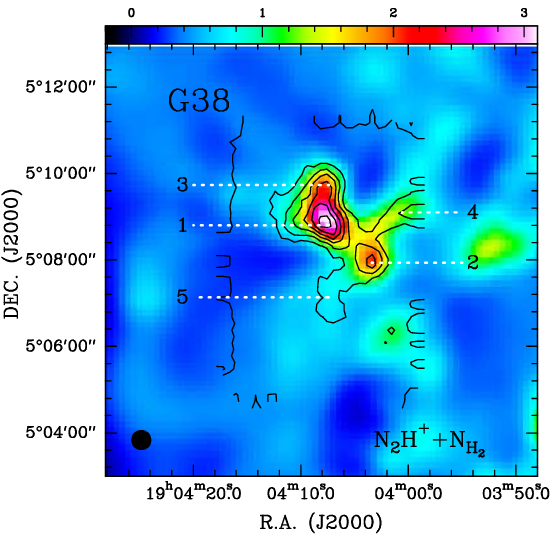}
\includegraphics[width=0.365\textwidth, angle=0]{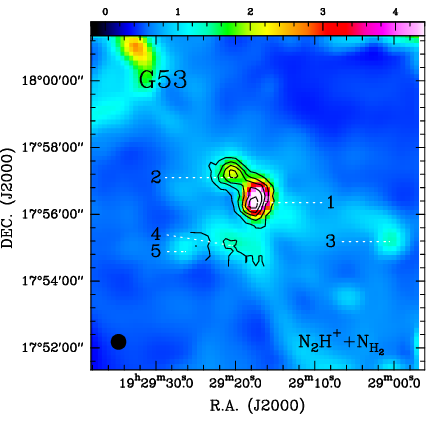}
\caption{The distributions of column density ($N_{\rm H_2}$) in color scale derived from SED fits pixel by pixel combining with \textit{Herschel} 70, 160, 250, 350, and 500 $\mu$m. The contours show the corresponding integrated intensities of N$_2$H$^+$ emission. The numbers indicate the positions of the selected cores, which are the same as those in \citet{Rathborne2006}. Contour levels and beam sizes are the same as those in Fig. \ref{Fig:int-map}. The unit of each color bar is in 10$^{22}$ cm$^{-2}$.}
\label{Fig:column_density}
\end{figure*}

The emissions from 8.0 $\mu$m to 70 $\mu$m are bright and extended at the northern part of G31, while they are weak at the southern part. We suggest that the filament nebula G31 consists of the southern IRDC and the northern infrared bright cloud (IRBC), as we contrast the consistency between their velocity distributions and the morphological distributions. It is obvious that the northern IRBC is more evolved than the southern IRDC. In the IRDC G33, the spectral observations covered only the northern part \citep{Rathborne2006}, however, the dark infrared background shows that this molecular cloud is still a filamentary structure in the north to south direction. The G33 has no extended infrared structure, and the infrared emissions at 8.0, 24, and 70 $\mu$m are relatively weak. The IRDC G34 is a well-known object \citep{Rathborne2005,Rathborne2008,Rathborne2011,Foster2014,Yanagida2014,Dirienzo2015,Pon2015,Pon2016,Xu2016}. From 24 $\mu$m to 870 $\mu$m, the three extracted dense cores in the G34 are very bright (see Fig. \ref{Fig:int-map}). At 8 $\mu$m, the cores G34-1 and G34-2 seem to be invisible. The G34-1, G34-2, and G34-3 have a dust temperature with 16.45, 20.02 and 19.49 K, respectively. A heating bridge exists, cutting through the cores G34-2 and G34-3 (see Fig. \ref{Fig:temperature}). The G34-1 and G34-2 are two of the most massive and dense cores within all the IRDCs. The IRDC G35 is also a well-studied object \citep{Henshaw2013,Henshaw2014,Jimenez2014,Dirienzo2015}, and has similar conditions to G35. The G35 is located in the W48 molecular complex, and has a very dark infrared emission observed by \textit{Spitzer}. The IRDC G38 is located between the \HII regions G38.91-0.44 and G39.30-1.04. The formation of this IRDC may be associated with the interaction of the two adjacent \HII regions \citep{Xu2013}, which suggests that the dense filamentary clumps may be formed during cloud-cloud collisions \citep{Tan2000,Tasker2009}. At 8.0 $\mu$m in the center of the core G53-2, the IRDC G53 presents a radial structure toward its surrounding. However, its absorption distribution at 8.0 $\mu$m shows an elongated structure from north to south. Therefore, the filamentary structure is a widespread phenomenon on a large scale \citep{Li2013,Wang2015}. The embedded dense cores within IRDCs and the \HII regions nearby influence the morphology of the host clouds.

An IRDC is a cold, dense region of a giant molecular cloud. They can be seen in silhouette against the bright diffuse infrared emission from the galactic plane \citep{Butler2009}. Observations in molecular clouds reveal a complex structure, with gas and dust often arranged in filamentary, rather than spherical geometries \citep[e.g.,][]{Hatchell2005,Myers2009}. Generally, in the line of sight, an IRDC presents filamentary structure with dark infrared wavelength or bright millimeter emission, for example, IRDCs G31, G33, G34, and G35 in Fig. \ref{Fig:int-map}. Whereas some of the IRDCs show radiated morphology, for example, IRDC SDC13 \citep{Peretto2014} and single spot, for example, IRDCs G38 and G53 in Fig. \ref{Fig:int-map}.

\subsection{Star formation and evolution}

\subsubsection{The associated high-mass star formation}

\begin{figure*}
\centering
\includegraphics[width=0.49\textwidth, angle=0]{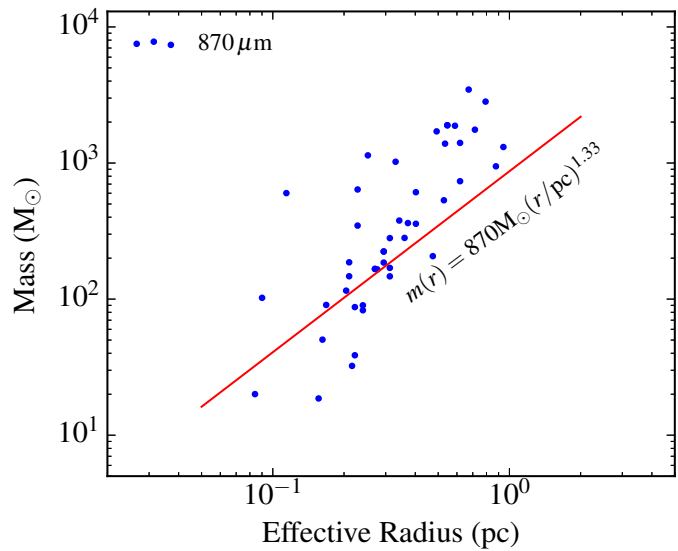}
\caption{Mass-radius distributions of Gaussian cores extracted from $Gaussclumps$. The masses are listed in Table \ref{tab:hcop}, and the effective radius is derived with $r = R_{\rm core} / (\sqrt{\rm ln2})$. The red line delineates the threshold introduced by \citet{Kauffmann2010} separating the regimes under which high-mass stars can form (above) or not (below line). The green points belong to the derived core masses from 870 $\mu$m. }
\label{Fig:mass_size}
\end{figure*}

In Fig. \ref{Fig:mass_size}, we present the mass-size relation diagram for the extracted cores at 870 $\mu$m. Comparison with the high-mass star formation threshold of $m(r) > 870 {\rm \Msun} (r/{\rm pc})^{1.33}$ empirically proposed by \citet{Kauffmann2010} allows us to determine whether these cores are capable of giving birth to massive stars. The data points are distributed above the threshold (shown in Fig. \ref{Fig:mass_size}) that discriminates between high and low mass star formation whose entries fall above and below the line, respectively, indicative of high-mass star-forming candidates. It appears that the most cores are high-mass star-forming candidates.

Virial parameter $\alpha_{\rm vir}$ can be used for judgment on whether cores are in a state of being gravitationally bound. In Table \ref{tab:n2hp}, the derived virial parameters $\alpha_{\rm vir}$ are mostly $\lesssim$ 1.0. The cores are gravitationally bound, potentially unstable and may collapse. In addition, these cores have high masses, densities, and low temperatures; and the star formation is active. Therefore, they are good high-mass star forming candidates in early stage.

\subsubsection{Evolutionary stages}

The derived dust temperature and column density distributions are the most direct evidence in presenting their evolutionary stages. In Figs. \ref{Fig:temperature} and \ref{Fig:column_density}, we can see that the temperature and density distributions have strong associations with the contours of $\nhp$, which is one good tracer in low temperature and high density. The active protostellar cores typically have warmer dust temperatures than the more quiescent, perhaps pre-protostellar cores \citep{Rathborne2010}. Generally in one star-forming region, these relatively high temperature spots are more evolved than these low temperature ones.

Infrared continuum emission is often used as a tracer in different evolutionary stages \citep{Churchwell2006,N131}. Mid-infrared continuum emission, for example, 24 $\mu$m, is an important tracer for test this evolutionary sequence. The IRDCs range from near-infrared dark to millimeter bright. The far-infrared 70 $\mu$m seems to be the turning point from infrared dark to infrared bright (see Fig. \ref{Fig:int-map}). The dense dust cores in this work basically have strong 870 $\mu$m and $\nhp$ emission, so they should be early star formation candidates. The infrared dark cores will be much earlier candidates than the infrared bright cores. For example in IRDC G38, the core G38-2 is probably a strong \HII region, where a protostar in late stage might reside. The G38-2 is apparently more evolved than the other cores in G38. The peak positions of some dense cores derived from N$_2$H$^+$ are offset from the ones with infrared emission, some of which indicates the site of \HII region. It is likely that the strong \HII regions in the late stage are triggering a new generation of star formation in the early stage.

\subsection{Dynamical characters}

\subsubsection{Outflow and infall}

In Fig. \ref{Fig:pv}, we show the position-velocity diagrams along the broken line (from the upper to lower) of each IRDC in Fig. \ref{Fig:int-map}. The diagrams can roughly present the positions with active dynamics, such as outflow and rotation. $\hcop$ is often optically thick, while $\nhp$ is optically thin, so both can be combined to use as dynamical tracers. Considering the line wings of $\hcop$ spectra in Fig. \ref{Fig:iram-spectra} and the position-velocity diagrams in Fig. \ref{Fig:pv}, we find that the potential outflow movement is ubiquitous in IRDCs, such as the 18 cores G31-1, G31-2, G31-4, G31-5, G31-6, G31-7, G31-8, G31-11, G34-1, G34-2, G34-4, G34-5, G35-6, G35-7, G35-9, G38-1, G38-3, and G53-1. We only detected SiO emission toward the cores G31-1, G34-1, G34-2, G34-3, G38-1, and G53-1 (see Fig. \ref{Fig:cso-spectra}). Considering that the molecule SiO is a good tracer of shock gas associated with molecular outflows \citep{Schilke1997}, these six sources are possible outflow candidates. The cores G31-4, G34-1, G34-2, G34-3, and G53-2 have SiO line widths ranging from 9.46 $\kms$ to 17.49 $\kms$, which further support this idea. However, G38-1 has a relatively narrow SiO line (5.01 $\kms$), its line-center velocity is about 2.5 $\kms$ offset from that of DCO$^+$. We note that there is a weak redshifted absorption dip. Based on this, it seems to be an inverse P-Cygni profile indicating a potential large scale infall or collapse there.

On the other hand, following the method of \citet{Mardones1997}, we can identify possible global infall motions of these cores using optically thick $\hcop$ and optically thin $\nhp$ lines. The blueshifted asymmetric $\hcop$ profiles with respect to that of $\nhp$ suggest that the ten cores G31-1, G31-2, G31-3, G33-1, G34-3, G35-6, G35-8, G38-2, G53-1, and G53-2 are possible global infall candidates. Many sources show both infall and outflow motions, as is consistent with the star-formation theory in \citet{Tan2014}. It is interesting to find that the optically thin $\co$ in the core G35-7 shows at least three velocity components, while the optically thick $\hcop$ shows broadening line width with blueshifted profile (see Fig. \ref{Fig:iram-spectra}). For G35 it has been suggested by \citet{Jimenez2014} that the three filamentary velocity components are converging in together, and the core G35-7 is the production in this converging flow. Our multi-velocity components detected in $\co$ seem to support their idea. However, it is somewhat difficult to give a final result as to whether it is infall movement or converging flow for the moment.

\subsubsection{Envelope expansion with core collapse}

The derived virial parameters $\alpha_{\rm vir}$ listed in Table \ref{tab:n2hp} are mostly $\lesssim$ 1.0, indicating that these cores are probably collapsing inward under the self-gravity. Additionally, it is interesting that the observed $\hcop$ spectra shown in Fig. \ref{Fig:iram-spectra} are blueshifted with stronger blueshifted and weaker redshifted peaks, while the spectra $\co$ are redshifted for the same core, for example, in cores G31-2, G31-3, G31-4, G31-6, G31-8, G31-10, G33-1, G33-4, G33-11, G35-7, and G35-8. In critical density, $\hcop$ can trace much denser regions of a core than $\co$. Generally, the innermost part of a core is denser than its envelope. Therefore, the $\hcop$ traces the inner dynamical feature within a core, while the $\co$ traces relatively the outer dynamical movement. Considering that the observed blueshifted and redshifted profiles for $\hcop$ and $\co$ in the same core, respectively, maybe that different parts of the core are moving in opposite directions, that is, inward versus outward. Such coexistence of inward and outward motions in star-forming molecular clouds or cores may be the ``envelope expansion with core collapse (EECC)'' \citep{Gao2010,Lou2011}. The clouds or cores B68 \citep{Lada2003}, SFO 11NE \citep{Thompson2004}, L1495A-N, L1507A, L1512 \citep{Lee2001}, L1517B \citep{Fu2011}, and L1689-SMM16 \citep{Chitsazzadeh2014} were also shown relative to the reports.

\subsection{Chemistry}

\subsubsection{Different tracers of star formation}

\begin{figure*}
\centering
\includegraphics[width=0.75\textwidth, angle=0]{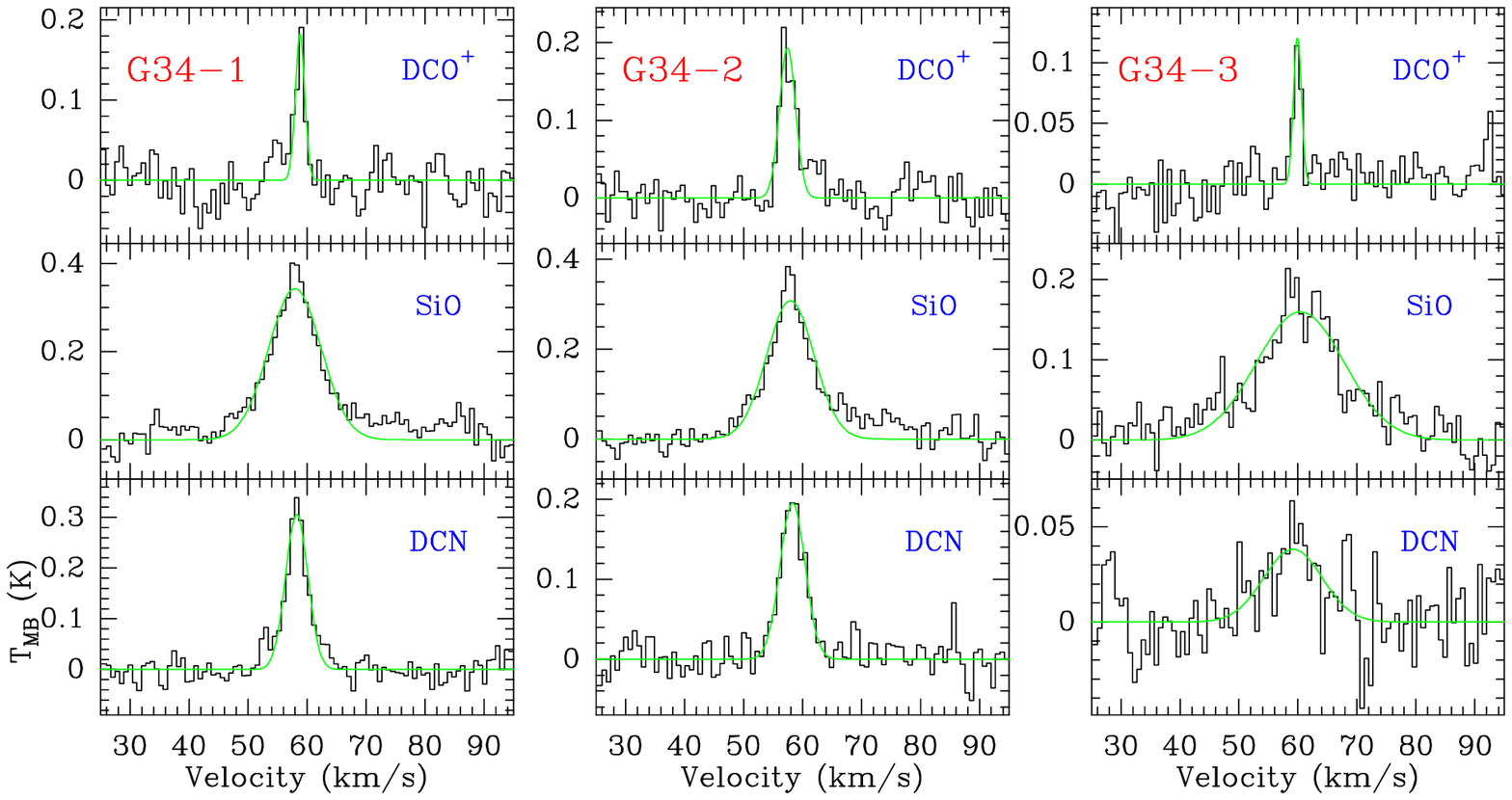}
\includegraphics[width=0.75\textwidth, angle=0]{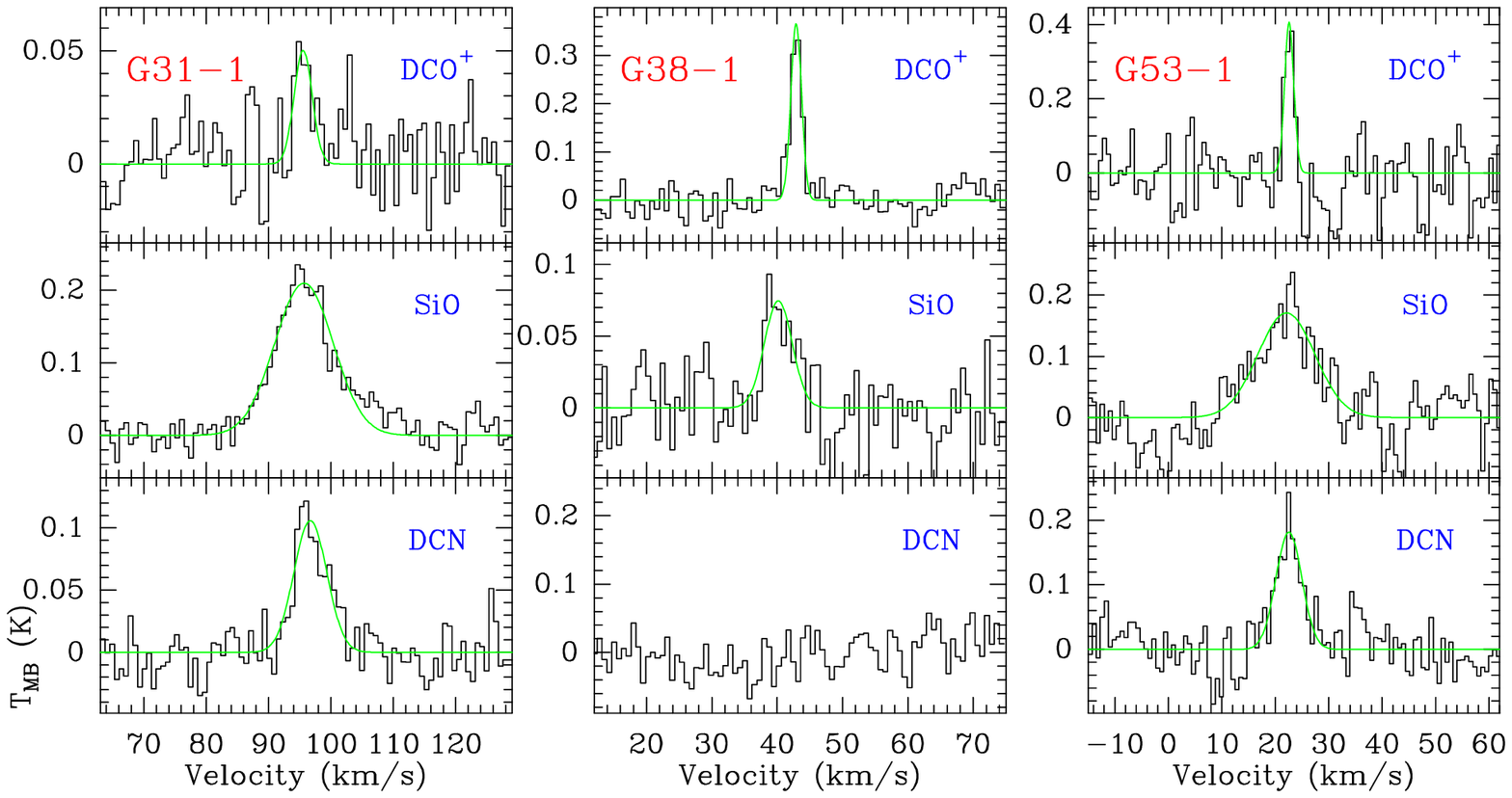}
\caption{Spectra of DCO$^+$, SiO, and DCN detected in cores G34-1, G34-2, G34-3, G31-1, G38-1, and G53-1. The green curves are from the Gaussian fit with GILDAS/CLASS software. The fitting parameters are listed in Table \ref{tab_cso-spectra}.}
\label{Fig:cso-spectra}
\end{figure*}

During the evolution of early star formation, plenty of spectra were produced, such as $\hcop$, HNC, $\nhp$, and $\co$ in Fig. \ref{Fig:iram-spectra}, and DCO$^+$, SiO, and DCN in Fig. \ref{Fig:cso-spectra}. Each spectral tracer has a special function. The $\co$ line has strong depletion toward the dense and cold cores \citep{Tafalla2004,Fontani2006,Hernandez2011a}, so $\co$ may indicate an evolved stage. These N- and D- bearing species have high critical density (> $10^5$ cm$^{-3}$), and can trace very low temperature (< 20 K) conditions \citep{Caselli1999,Tafalla2002,Bergin2007}. SiO is a powerful probe of star formation activity in the deeply embedded phase of the evolution of massive cores \citep{Csengeri2016}, and a characteristic tracer of shocks \citep{Jimenez2010,Hirano2010}.

We tried to observe deuterium species in all the potentially dense cores in Table \ref{tab:n2hp}. However, we detected only six DCO$^+$ and five DCN lines which are mainly located at cores G31-1, G34-1, G34-2, G34-3, G38-1, and G53-1 in Fig. \ref{Fig:cso-spectra} and Table \ref{tab_cso-spectra}. Likely due to relatively small dynamical motions, the deuterium species, especially for DCO$^+$, have narrower line width than these non-D-bearing species. They are suggested as tracers of the dense and quiescent part of cold cores \citep{Bergin2007,Francesco2007}. \citet{Pagani2011} argued that the absence of DCO$^+$ outside depleted cores is due to a high ortho-H$_{2}$ abundance in dark clouds maintained long enough. Furthermore, it is difficult to detect deuterium species, such as DCO$^+$ and DCN. One of the reasons for this is that their low abundance. We also argue that the condition producing such deuterium species restricts in a small area relatively to beam size of telescope, so there exists high beam dilution. The high spatial resolution observation is necessary to carry out this investigation.

With 870 $\mu$m continuum, we derived higher densities and masses than the work by \citet{Liu2014}, who mainly used $^{13}$CO and C$^{18}$O emission in studying IRDCs. It is suggested that both the CO species are bad tracers in the condition of IRDCs. The CO lines, at least, are not suitable for computing the density and mass of these dense cores. The N- and D- bearing species and submillimeter continuum are better tracers in investigating the cold and dense IRDCs.

\subsubsection{Molecular depletion}

In Fig. \ref{Fig:int-map}, we found that the morphology of $\co$ distribution is very different from those of $\nhp$ and dust emission (i.e., G33-1, G34-1, G34-3, G35-4, and G35-7). Especially, we can see by eye that the peaks of $\co$ emission present offsets from the peak positions of those cold and dense cores. It is likely that $\co$ emission is heavily depleted due to freezing out onto dust grains in cold and dense condition. Their $\co$ lines in Fig. \ref{Fig:iram-spectra} also show a non-Gaussian profile, further proving that the $\co$ emission in the cores was depleted onto the cold grains nearby. It is also possible that the chemical components (not only for $\co$) should be affected in the cold and dense conditions.

In Fig. \ref{Fig:N-int}, the relationship between column density and molecular integrated intensity is very clear. The best correlation is the relationship between column density and $\nhp$ integrated intensity; the next-best correlation is with HNC and then $\hcop$; the worst correlation is with $\co$. This suggests that $\nhp$ is a better tracer in dense conditions of column density > $\sim$10$^{22}$ cm$^{-2}$ than HNC, $\hcop$, and $\co$. In Fig. \ref{Fig:int-int}, we show the relationship between integrated intensities between $\hcop$, HNC, $\nhp$, and $\co$. The best correlation is the one between HNC and $\nhp$, while the worst one is between $\co$ and $\nhp$. In dense conditions, $\hcop$ and $\co$ may become optically thick \citep{Pillai2006}. $\co$ has a relatively low critical density and is vulnerable to freeze out onto the surface of cold dust grains. Therefore, $\nhp$ is also a good tracer in very dense and cold (below 20 K) condition, where $\co$ is a unreliable tracer.

\begin{figure*}
\centering
\includegraphics[width=0.33\textwidth, angle=0]{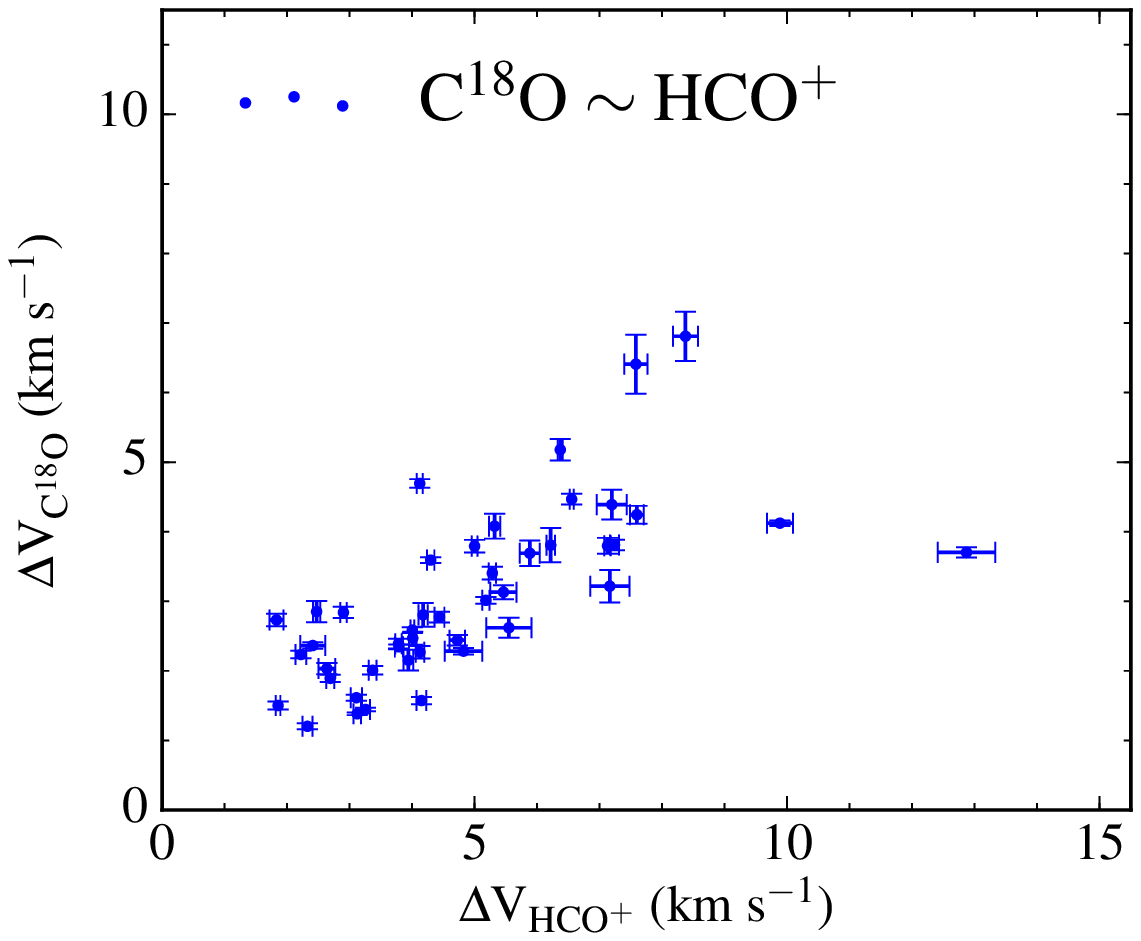}
\includegraphics[width=0.33\textwidth, angle=0]{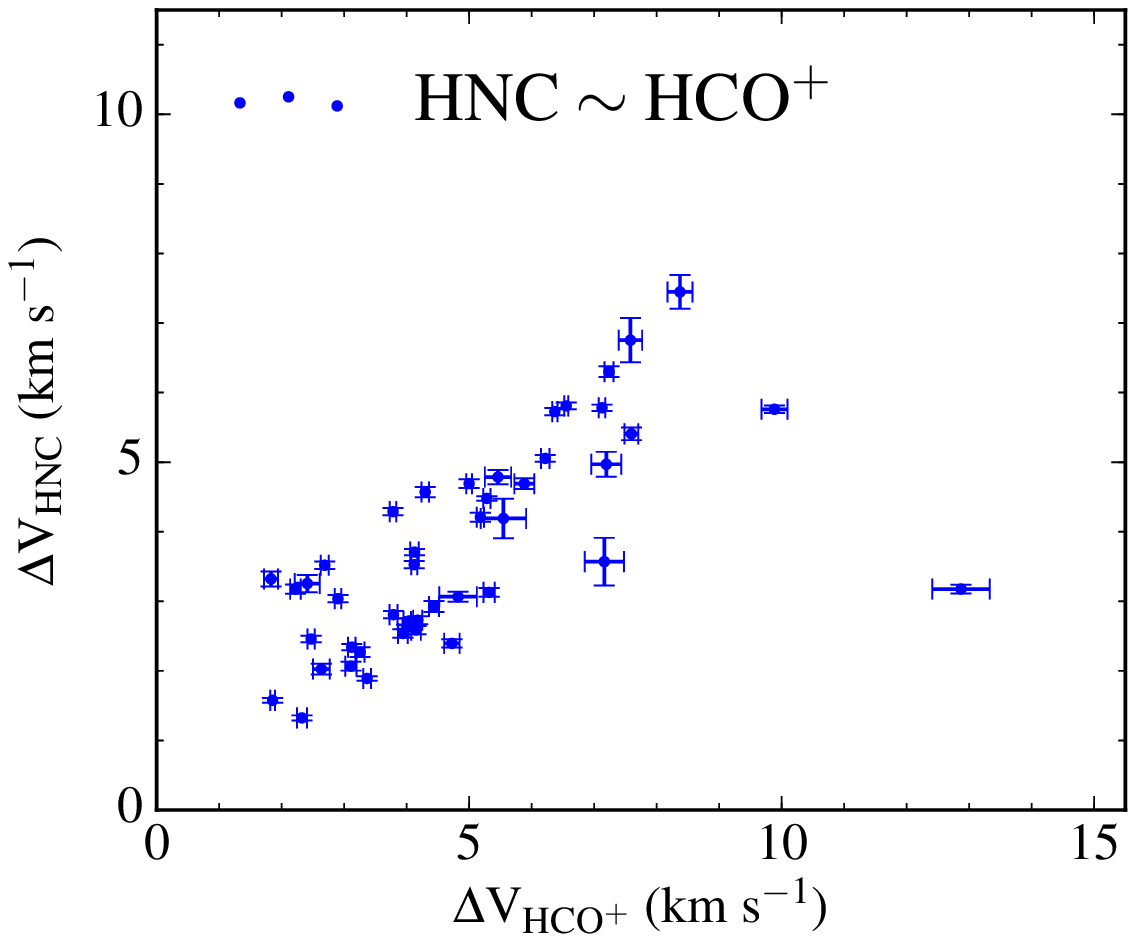}
\includegraphics[width=0.33\textwidth, angle=0]{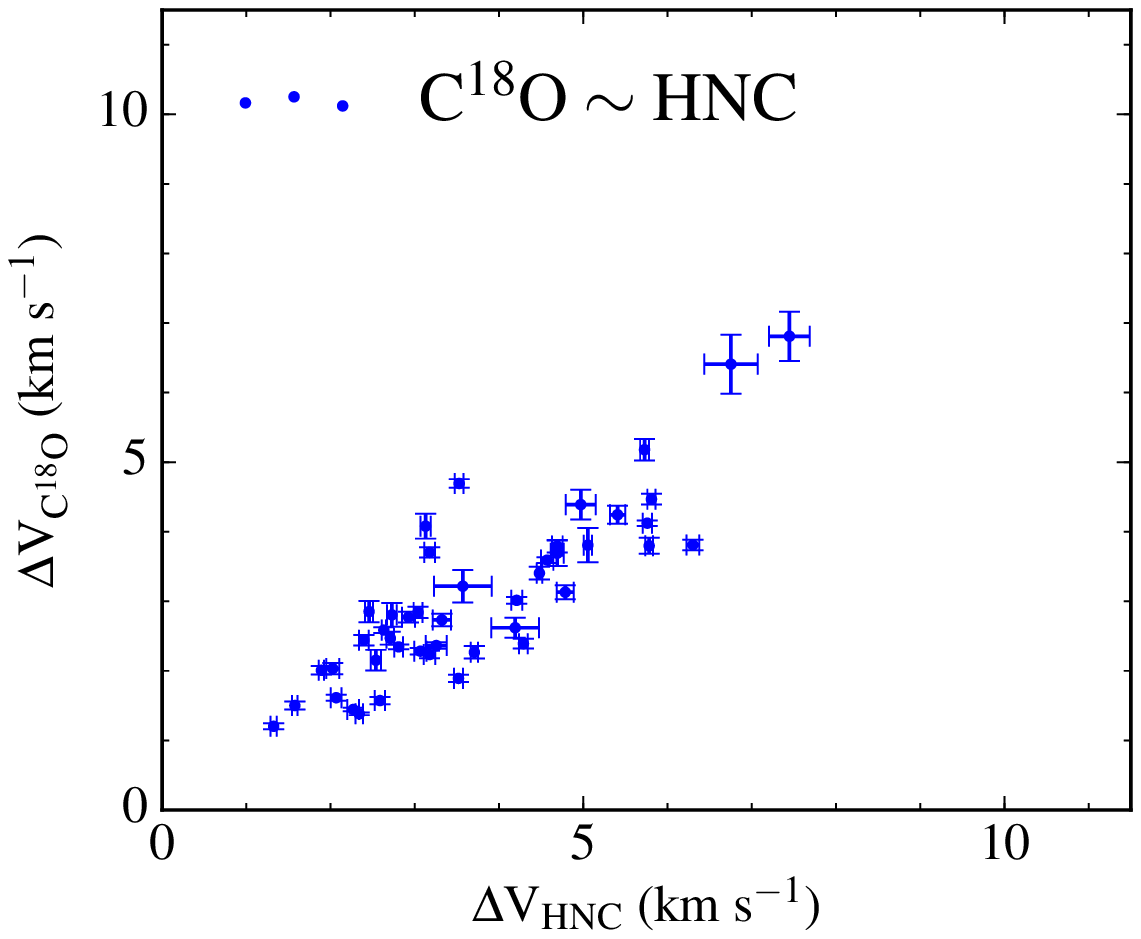}
\caption{Line width ($\Delta V$) diagrams between $\hcop$, HNC, and $\co$. The data points are listed the Table \ref{tab:hcop}.}
\label{Fig:width}
\end{figure*}

\begin{figure*}
\centering
\includegraphics[width=0.33\textwidth, angle=0]{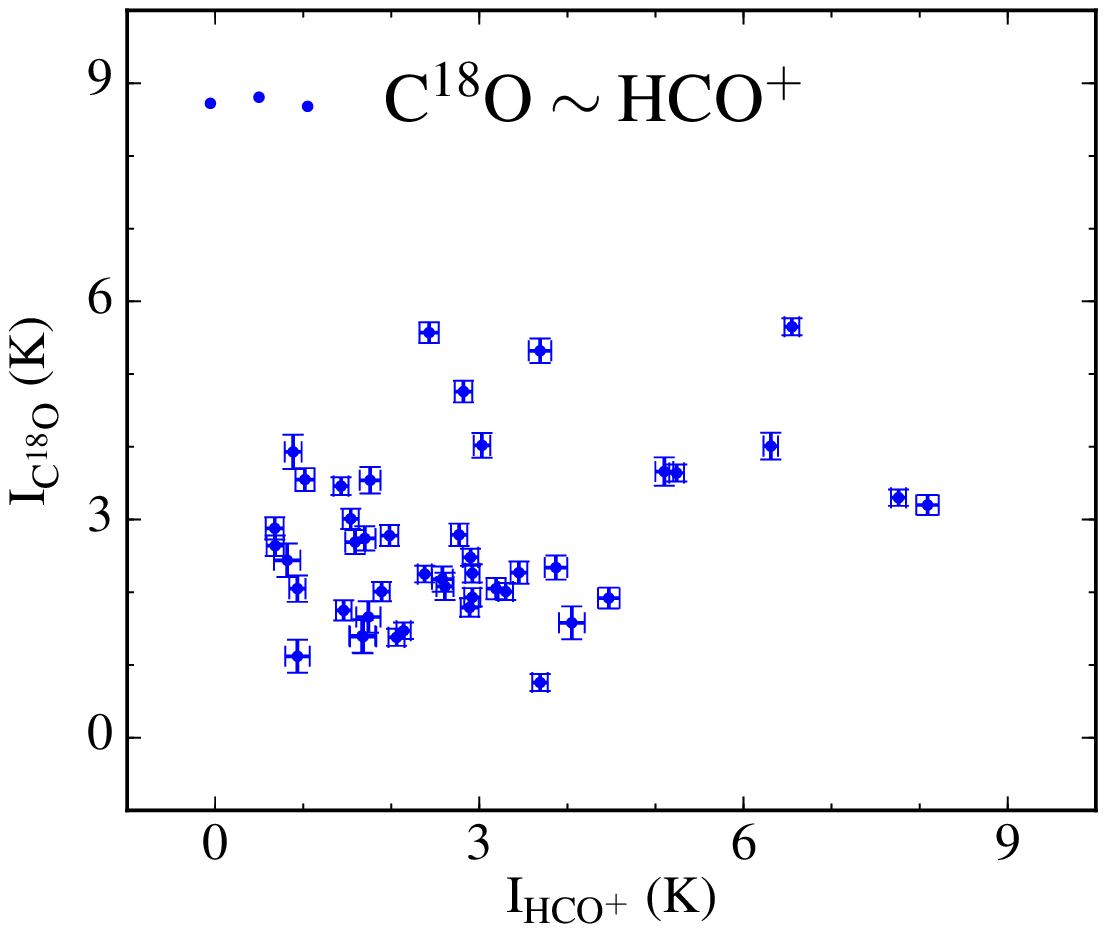}
\includegraphics[width=0.33\textwidth, angle=0]{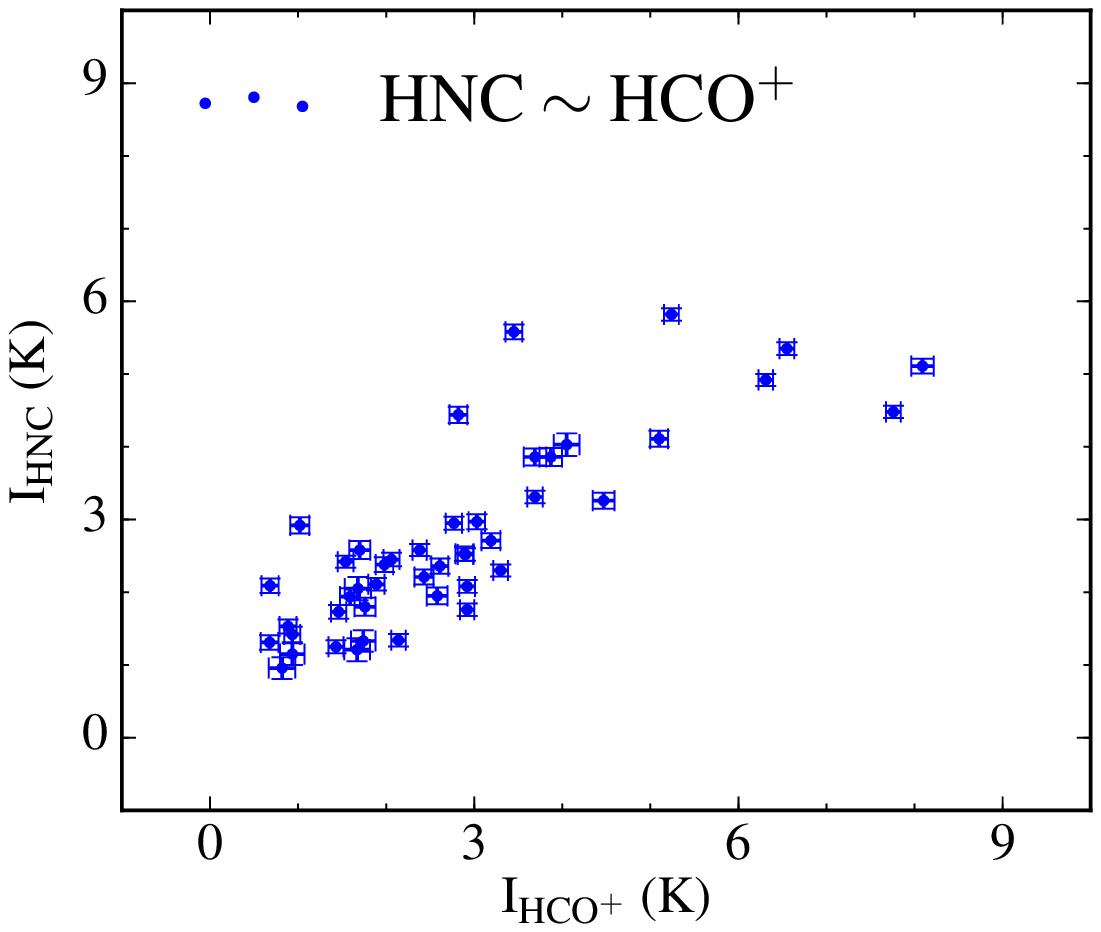}
\includegraphics[width=0.33\textwidth, angle=0]{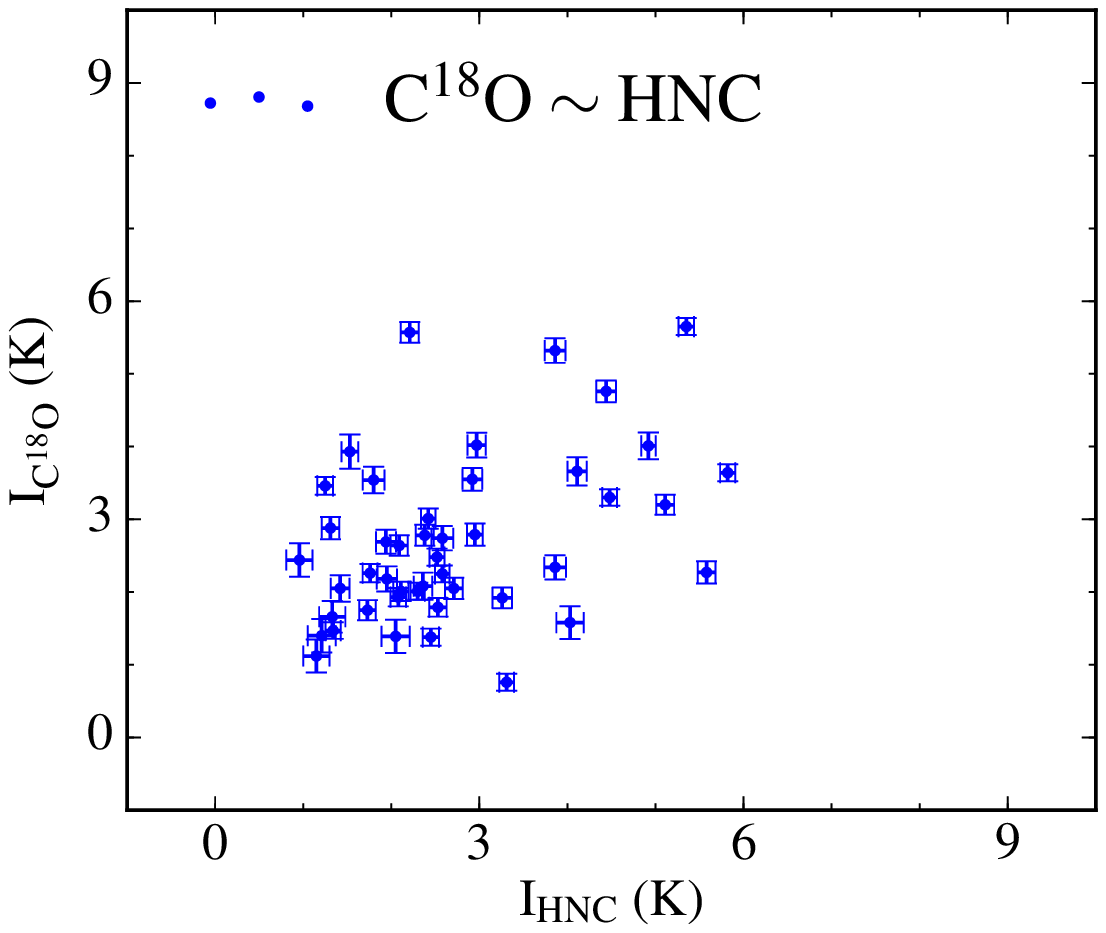}
\caption{Line intensity ($I$) diagrams between of $\hcop$, HNC, and $\co$, respectively. The data points are listed the Table \ref{tab:hcop}.}
\label{Fig:intensity}
\end{figure*}

Line width ($\Delta V$) and line intensity ($I$) diagrams are shown in Figs. \ref{Fig:width}
and \ref{Fig:intensity}, respectively, demonstrating the relations between all combinations of  $\hcop$, HNC, and $\co$. The parameters of line width and intensity are derived from Gaussian fits, masking the absorption dip (see Section \ref{sect:data_iram30}). Although the figures do not  show the correlation coefficient, it is obvious that the correlation of line width between $\co$ and HNC is better than that between HNC and $\hcop$, and than that between $\co$ and $\hcop$ in Fig. \ref{Fig:width}. This can be explained as the dynamical broadening has a greater effect on the spectra of $\hcop$ than of HNC and C$^{18}$O. It further proves that HCO$^+$ is an important dynamical tracer in cores. Additionally, the correlation of line intensity between HNC and $\hcop$ is better than that between $\co$ and HNC, and between $\co$ and $\hcop$ in Fig. \ref{Fig:intensity}. This can be explained that $\co$ has a relatively lower critical density than HNC and $\hcop$, while HNC and $\hcop$ has a similar critical density. It is also likely that $\co$ depletes more strongly than HNC and $\hcop$ in dense and cold cores.

\section{Summary}
\label{sect:summary}

We mainly used spectral data from the IRAM 30 m and CSO 10.4 m telescopes combining with continuum data from online archive to study the physical conditions in six IRDCs including G031.97+00.07, G033.69-00.01, G034.43+00.24, G035.39-00.33, G038.95-00.47, and G053.11+00.05. The newly observed spectral data are $\hcop$, HCN, $\nhp$, $\co$, DCO$^+$, SiO, and DCN, respectively.

We investigated 44 cores including 37 cores reported in \citet{Rathborne2006} and 7 newly-identified cores. Toward the dense and cold cores, we detected 6 DCO$^+$, and 5 DCN lines.
By fitting the seven HFS lines from the N$_2$H$^+$, we find that the optical depth mainly ranges from 0.10 to 0.24, consistent with online RADEX results, so the N$_2$H$^+$ is optically thin in the condition of these dense and cold cores.

Using pixel-by-pixel SED fits of the \textit{Herschel} 70 to 500 $\mu$m, we obtained the dust temperature and column density distributions of the IRDCs, and found that the IRDCs have a low dust temperature and a high column density. We also found that $\nhp$ emission has a strong linear correlation with the dust temperature and column density distributions.

The peak positions of some dense cores derived from N$_2$H$^+$ are offset from the ones with infrared emission, some of which indicate the site of the \HII region. It is likely that the strong \HII regions are triggering corresponding star formation around themselves.

We find that the morphologies of $\co$ distribution are significantly different from that of $\nhp$. It is obvious that $\co$ emission may be heavily depleted at the peak positions of these cold cores. Their $\co$ lines also show a non-Gaussian profile, further proving that the $\co$ emission in the cores was depleted onto the cold grains nearby. Comparing the relationships between column density and molecular integrated intensity, we found that the molecular ion $\nhp$ has the best correlation, while the worst is $\co$. It is suggested that $\nhp$ is a good tracer in very dense conditions, however a unreliable one is $\co$, because of its low critical density and vulnerability to freeze out onto the surface of cold dust grains.

We find that the infall, outflow, and collapse movements are active in IRDCs, suggesting active dynamics. In the investigated 44 cores, about 18(10) cores have potential outflow(infall) movements. For the observed blueshifted and redshifted profiles for $\hcop$ and $\co$ in the same core, respectively, maybe that different parts of the core are moving in opposite directions, that is, inward versus outward. Such coexistence of inward and outward motions in star-forming molecular clouds or cores may be the so called ``envelope expansion with core collapse (EECC)''.

\section*{Acknowledgements}
\addcontentsline{toc}{section}{Acknowledgements}

We wish to thank the anonymous referee for comments and suggestions that improved the clarity of the paper. We acknowledge observations by the IRAM and CSO staff. C.-P. Zhang is supported by the Young Researcher Grant of National Astronomical Observatories, Chinese Academy of Sciences. G.-X. Li is supported by the Deutsche Forschungsgemeinschaft (DFG) priority program 1573 ISM- SPP. This work is partly supported by the National Key Basic Research Program of China (973 Program) 2015CB857100 and National Natural Science Foundation of China 11363004, 11403042, 11503035, 11573036, and 11373062.

\bibliographystyle{aa}
\bibliography{references}

\onecolumn

\begin{figure*}
\centering
\includegraphics[width = 0.99 \textwidth]{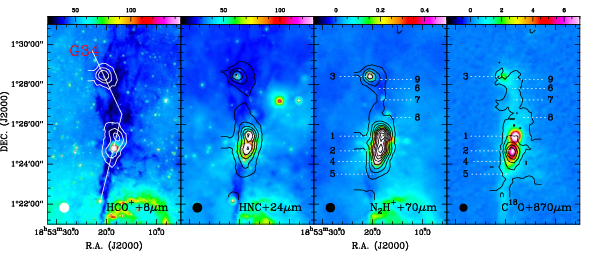}
\includegraphics[width = 0.99 \textwidth]{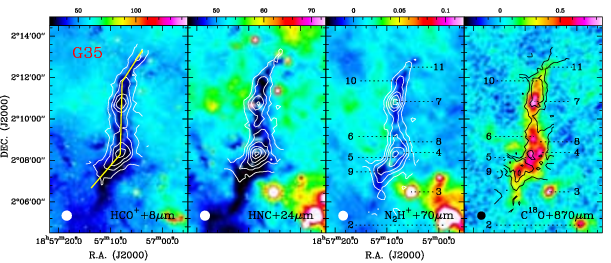}
\centerline{\textbf{Figure \ref{Fig:int-map}.} --- Continued.}
\end{figure*}

\begin{figure*}
\centering
\includegraphics[width = 0.64 \textwidth]{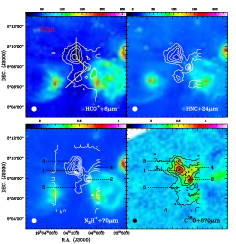}
\includegraphics[width = 0.64 \textwidth]{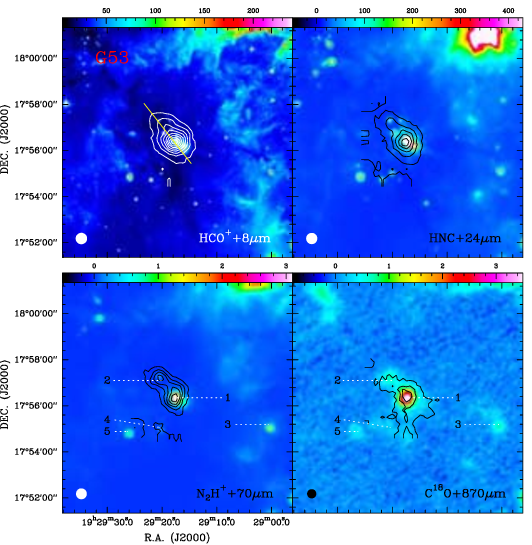}
\centerline{\textbf{Figure \ref{Fig:int-map}.} --- Continued.}
\end{figure*}

\begin{figure*}
\centering
\includegraphics[width=0.95\textwidth, angle=0]{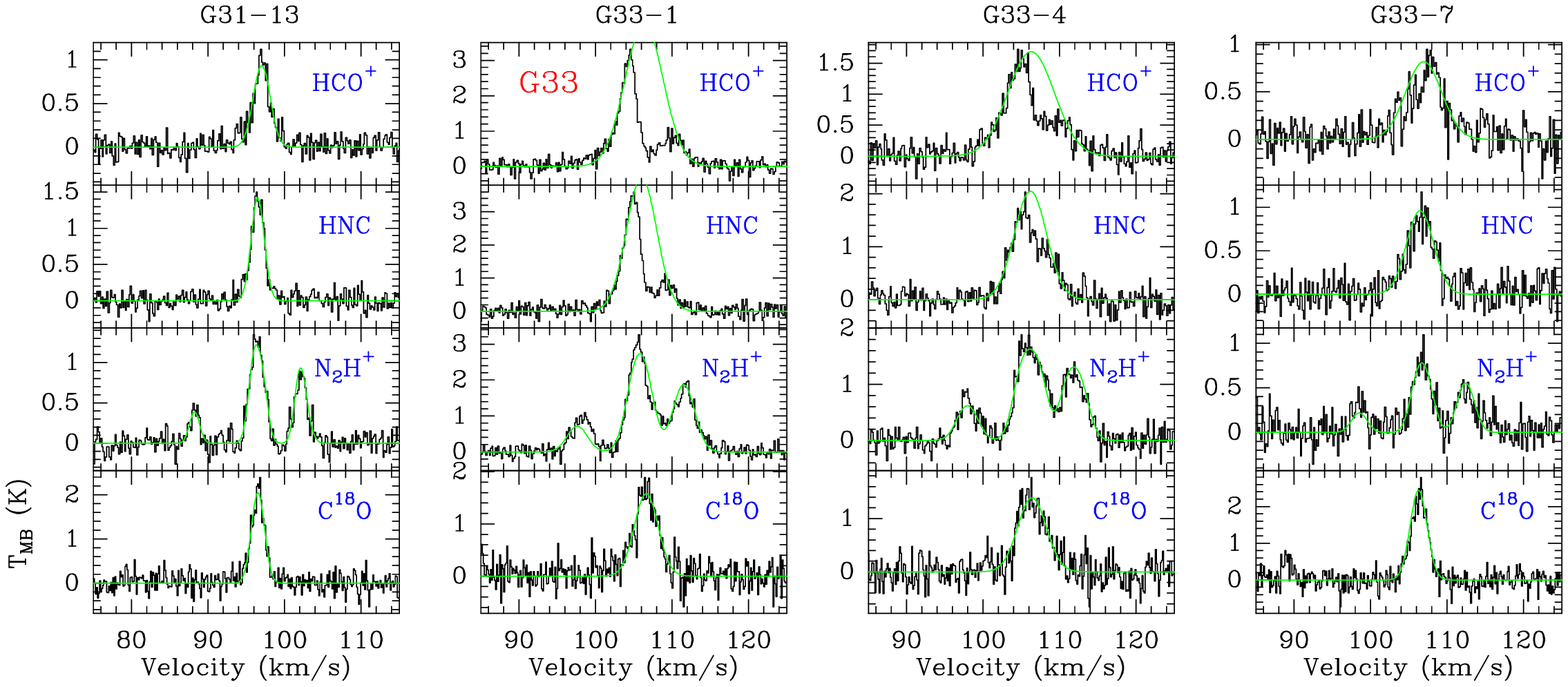}
\includegraphics[width=0.95\textwidth, angle=0]{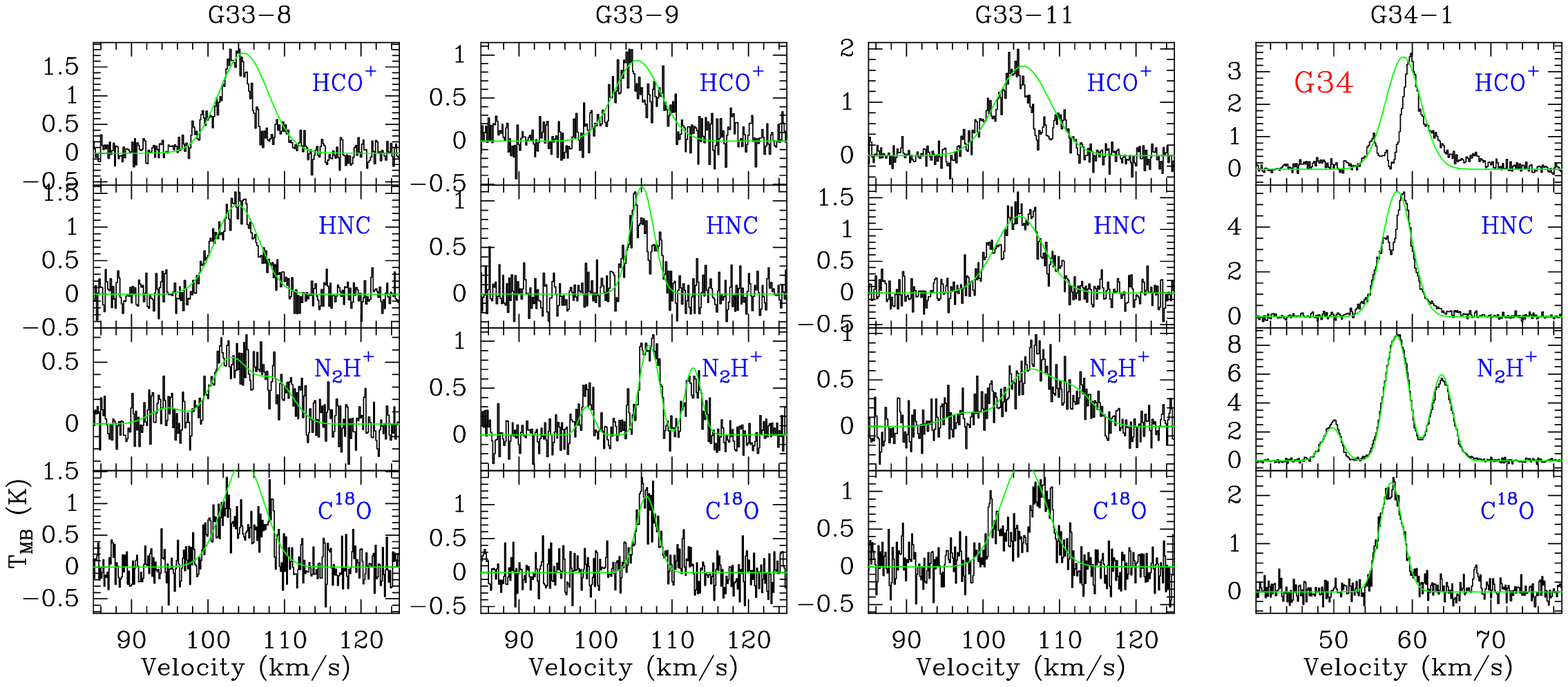}
\includegraphics[width=0.95\textwidth, angle=0]{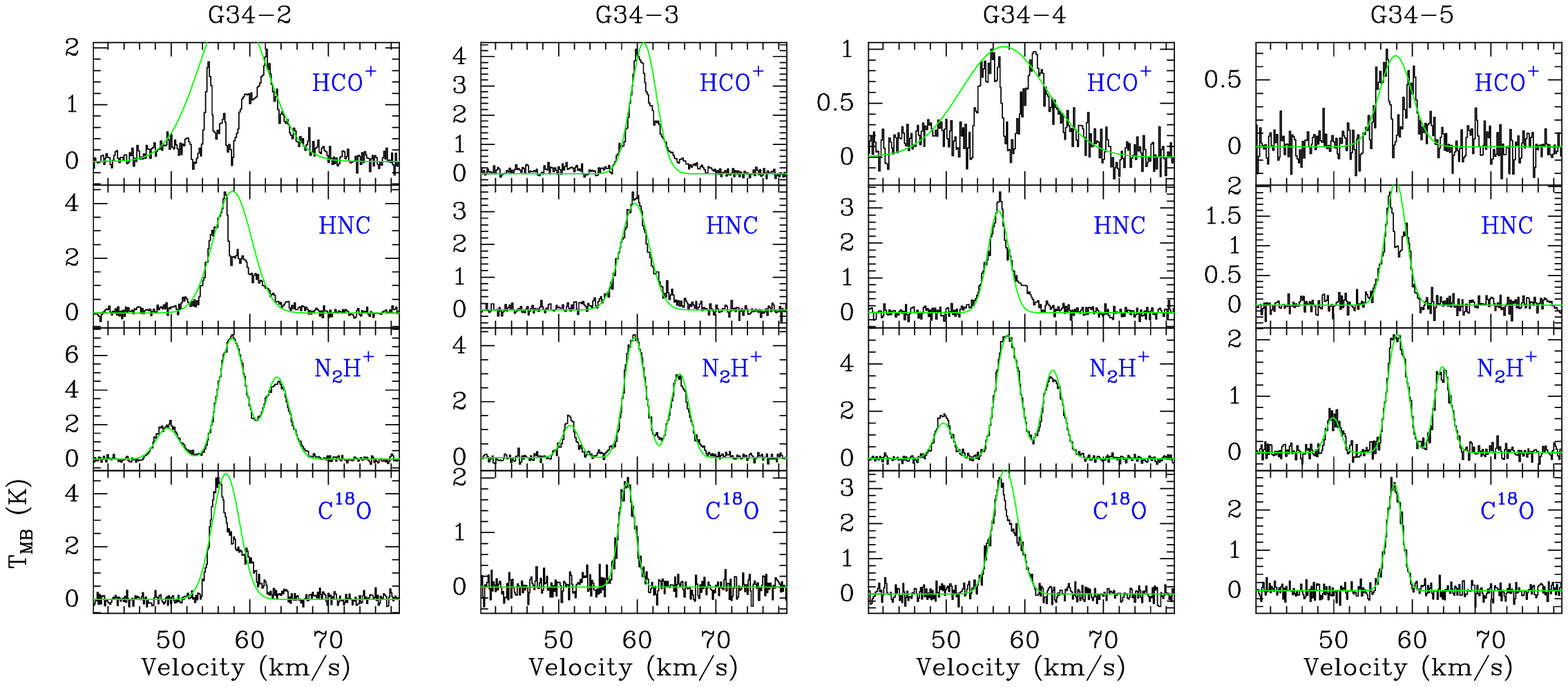}
\centerline{\textbf{Figure \ref{Fig:iram-spectra}.} --- Continued.}
\end{figure*}

\begin{figure*}
\centering
\includegraphics[width=0.95\textwidth, angle=0]{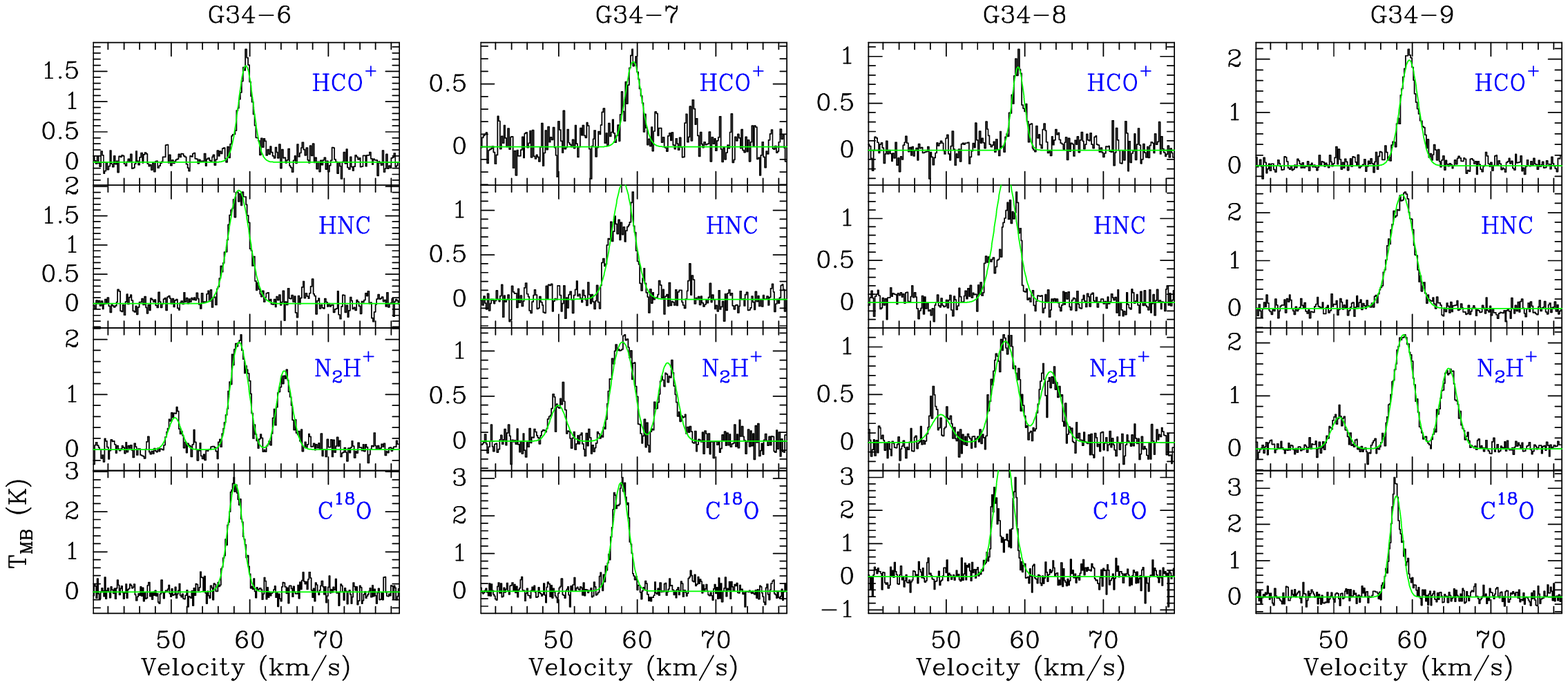}
\includegraphics[width=0.95\textwidth, angle=0]{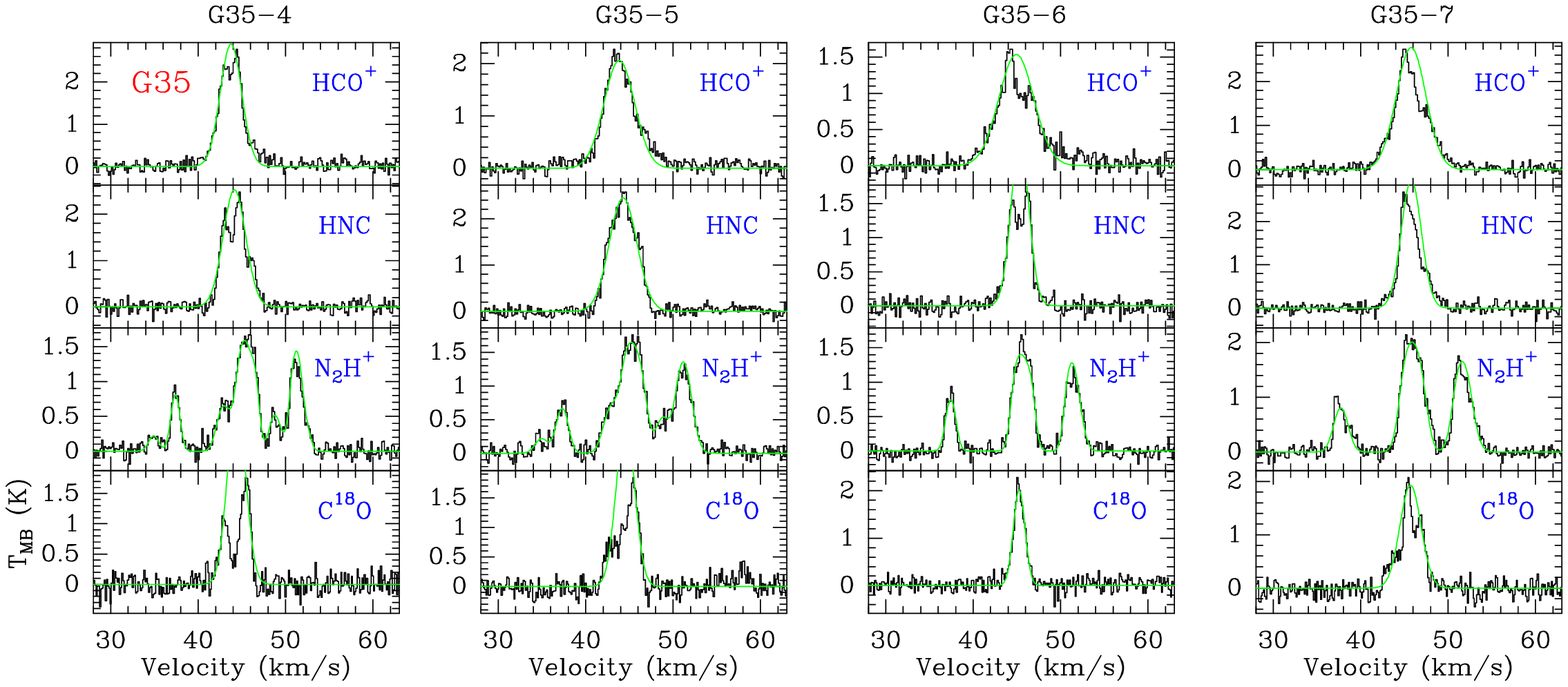}
\includegraphics[width=0.95\textwidth, angle=0]{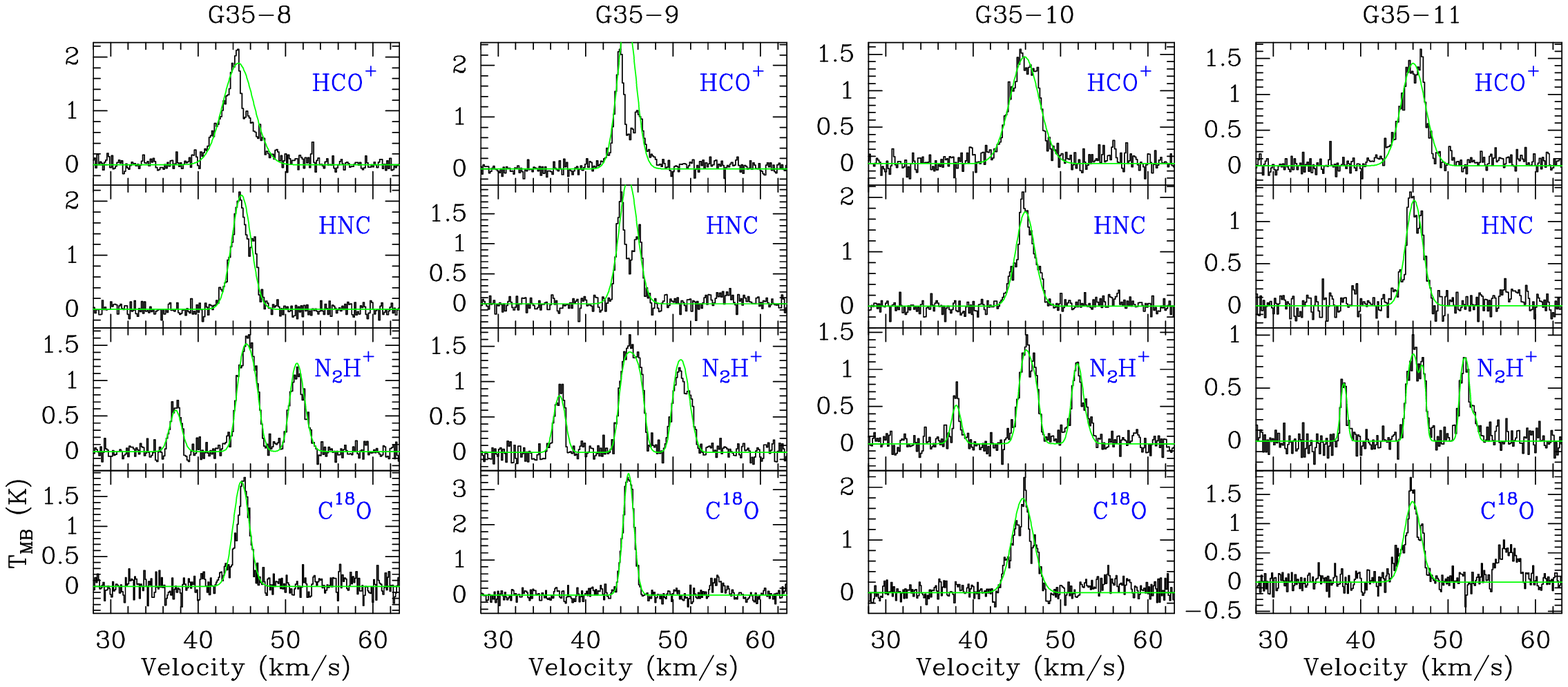}
\centerline{\textbf{Figure \ref{Fig:iram-spectra}.} --- Continued.}
\end{figure*}

\begin{figure*}
\centering
\includegraphics[width=0.95\textwidth, angle=0]{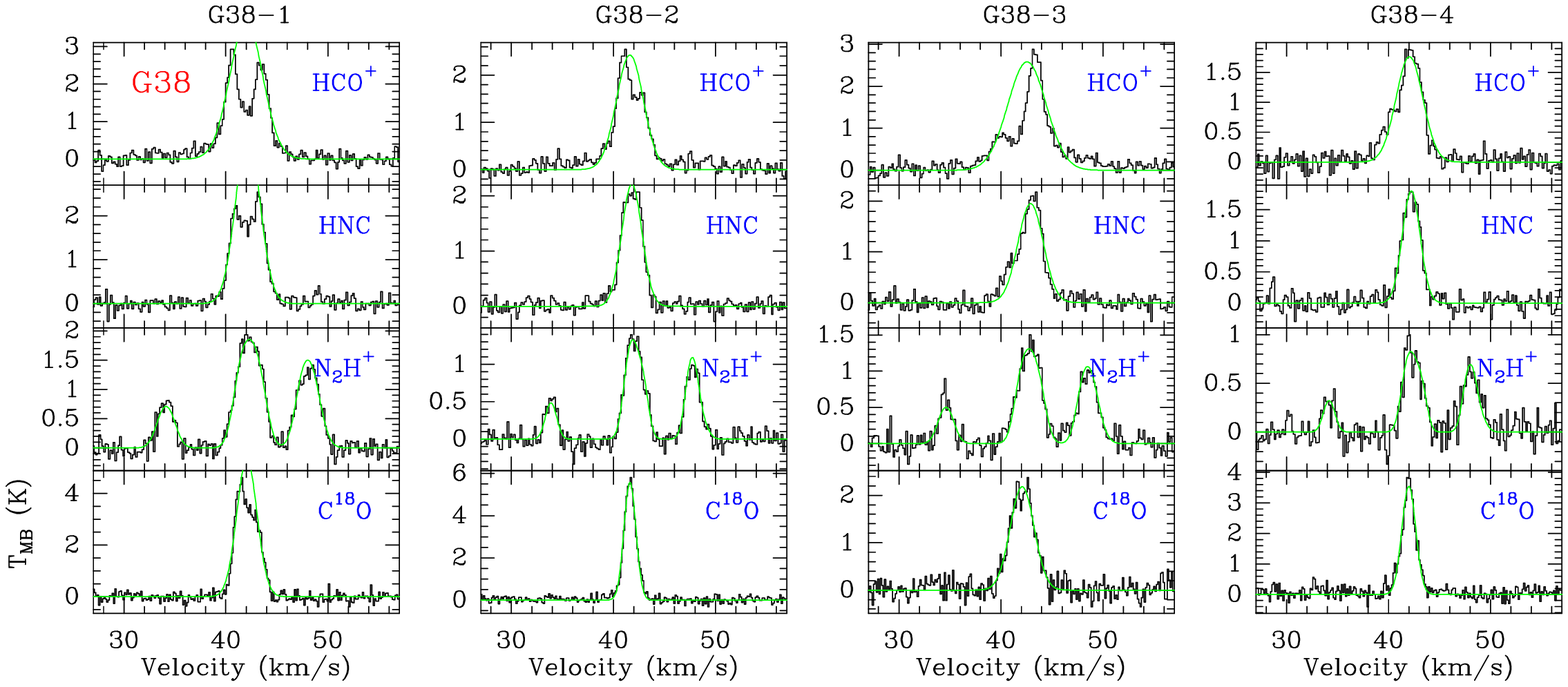}
\includegraphics[width=0.95\textwidth, angle=0]{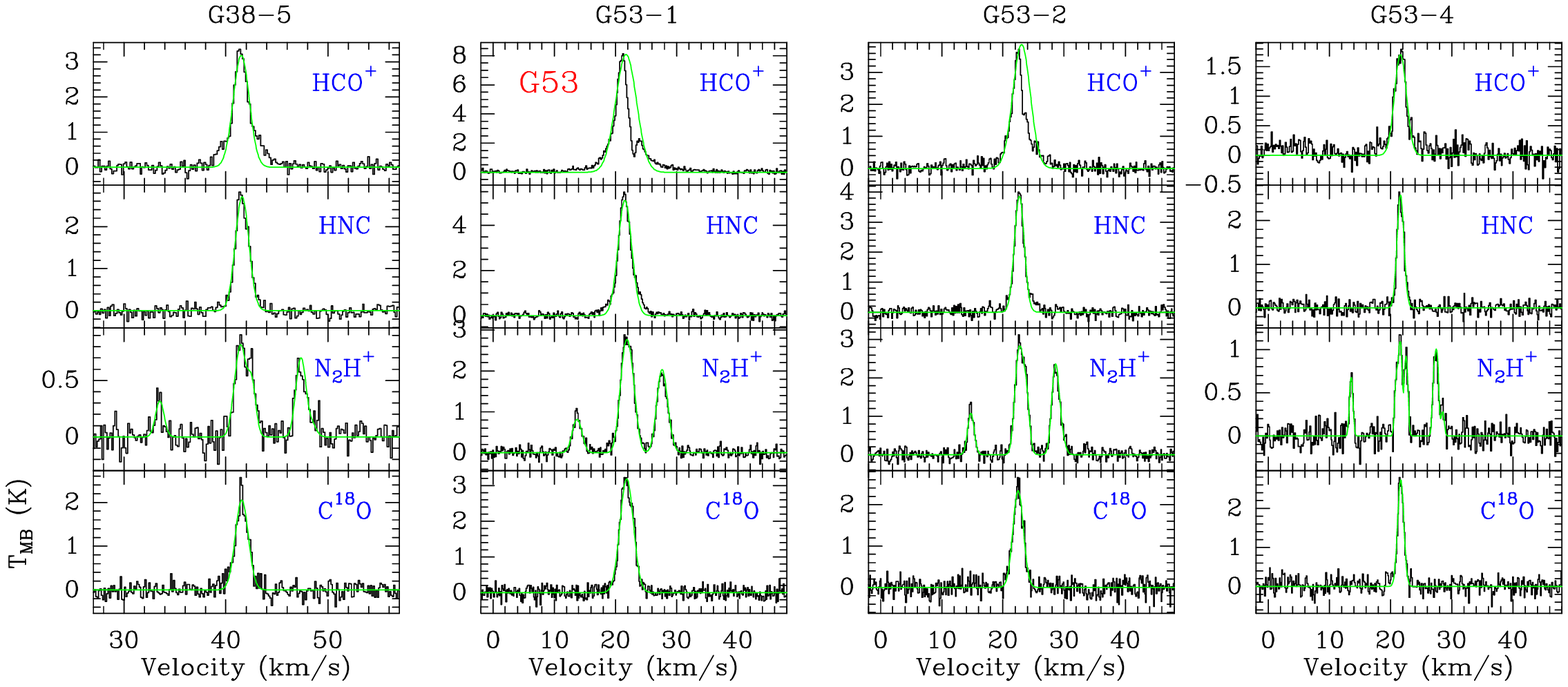}
\centerline{\textbf{Figure \ref{Fig:iram-spectra}.} --- Continued.}
\end{figure*}

\clearpage

\begin{table*}
\caption{Information of IRAM 30 m and CSO 10.4 m observations.}
\label{tab_obs} \centering 
\begin{tabular}{cccccc}
\hline \hline
Telescope & Observing date & Frequency &  Spectra  &  FWHM &  Mode  \\
 \hline
IRAM 30 m   & Dec. 2013  &  89 GHz --  &  $\hcop$, HNC, & 29.0$''$ -- & Mapping \\
           & Apr. 2014  &  110 GHz    &$\nhp$, $\co$   & 23.5$''$   &     \\
\hline
CSO 10.4 m  & Jul. 2015  &  216 GHz --  &  DCO$^+$,     & $\sim$33.4$''$   & Pointing\\
           &            &  218 GHz     &  SiO, DCN   \\
\hline
\end{tabular}
\end{table*}

\begin{landscape}
\begin{table*}
\caption{Parameters (coordinate, distance, mass, density, etc.) of the selected cores in the IRDCs.}
\label{tab:n2hp} \centering 
\footnotesize
\begin{tabular}{lcccccccccccccc}
\hline \hline
Core & R.A. & DEC. & Dist. & $R_{\rm core}$     & $T_{\rm dust}$ & $I_{\rm 870\, \mu m}$  & $S_{\rm 870\, \mu
m}$ & $\Delta V_{\rm N_2H^+}$ & $\tau_{\rm N_2H^+}$  &   $M_{\rm vir}$ &  $M_{\rm 870\, \mu m}$ & $M_{\rm 1.1 \, mm}$ &
$\alpha_{\rm vir}$& $n_{\rm 870\, \mu m}$   \\
 & J2000  & J2000 & kpc & pc &K&  $\jyb$&  Jy & $\kms$  & & $\Msun$ & $\Msun$ & $\Msun$ & & $\rm 10^4\, cm^{-3}$  \\
\hline
G31-1 & 18:49:36.3 & -00:45:45 & 6.9 & 0.20 &$ 19.3 \pm 0.2 $&$ 5.64 \pm 0.27 $&$ 4.42 \pm 0.21 $&$ 3.48 \pm 0.01 $&$ 0.12 \pm 0.00 $&$ 146 \pm 0 $&$ 1221 \pm 61 $&$ 1890 $&$ 0.12 \pm 0.01 $&$ 5.1 \pm 0.3 $ \\
G31-2 & 18:49:36.0 & -00:46:16 & 6.9 & 0.28 &$ 18.3 \pm 0.0 $&$ 3.16 \pm 0.15 $&$ 3.43 \pm 0.16 $&$ 3.10 \pm 0.01 $&$ 0.10 \pm 0.00 $&$ 179 \pm 1 $&$ 1022 \pm 47 $&$ 929 $&$ 0.18 \pm 0.01 $&$ 4.3 \pm 0.2 $ \\
G31-3 & 18:49:32.3 & -00:47:02 & 6.9 & 0.46 &$ 16.2 \pm 0.3 $&$ 2.14 \pm 0.08 $&$ 5.23 \pm 0.18 $&$ 2.19 \pm 0.52 $&$ 0.25 \pm 0.10 $&$ 209 \pm 50 $&$ 1895 \pm 86 $&$ 1222 $&$ 0.11 \pm 0.03 $&$ 7.9 \pm 0.4 $ \\
G31-3 & 18:49:32.3 & -00:47:02 & 6.9 & 0.46 &$ 16.2 \pm 0.3 $&$ 2.14 \pm 0.08 $&$ 5.23 \pm 0.18 $&$ 3.85 \pm 0.52 $&$ 0.53 \pm 0.10 $&$ 368 \pm 50 $&$ 1895 \pm 86 $&$ 1222 $&$ 0.19 \pm 0.03 $&$ 7.9 \pm 0.4 $ \\
G31-4 & 18:49:33.0 & -00:47:33 & 6.9 & 0.49 &$ 15.5 \pm 0.2 $&$ 1.74 \pm 0.03 $&$ 4.84 \pm 0.09 $&$ 3.71 \pm 0.04 $&$ 0.10 \pm 0.00 $&$ 381 \pm 4 $&$ 1879 \pm 49 $&$ 852 $&$ 0.20 \pm 0.01 $&$ 7.8 \pm 0.2 $ \\
G31-5 & 18:49:21.9 & -00:50:35 & 6.9 & 0.17 &$ 17.1 \pm 0.2 $&$ 0.64 \pm 0.00 $&$ 0.44 \pm 0.00 $&$ 2.74 \pm 0.04 $&$ 0.10 \pm 0.01 $&$ 101 \pm 2 $&$ 147 \pm 3 $&$ 178 $&$ 0.68 \pm 0.02 $&$ 0.6 \pm 0.0 $ \\
G31-6 & 18:49:35.0 & -00:46:44 & 6.9 & 0.66 &$ 17.2 \pm 0.4 $&$ 1.89 \pm 0.04 $&$ 8.60 \pm 0.16 $&$ 3.09 \pm 0.02 $&$ 0.10 \pm 0.00 $&$ 429 \pm 3 $&$ 2827 \pm 111 $&$ 1181 $&$ 0.15 \pm 0.01 $&$ 11.8 \pm 0.5 $ \\
G31-7 & 18:49:28.4 & -00:48:54 & 6.9 & 0.34 &$ 17.5 \pm 0.1 $&$ 0.78 \pm 0.00 $&$ 1.12 \pm 0.01 $&$ 1.97 \pm 0.07 $&$ 0.11 \pm 0.03 $&$ 138 \pm 5 $&$ 358 \pm 4 $&$ 291 $&$ 0.39 \pm 0.01 $&$ 1.5 \pm 0.0 $ \\
G31-8 & 18:49:29.1 & -00:48:12 & 6.9 & 0.52 &$ 17.6 \pm 0.2 $&$ 0.78 \pm 0.01 $&$ 2.32 \pm 0.02 $&$ 2.06 \pm 0.04 $&$ 0.10 \pm 0.00 $&$ 223 \pm 5 $&$ 735 \pm 16 $&$ 493 $&$ 0.30 \pm 0.01 $&$ 3.1 \pm 0.1 $ \\
G31-9 & 18:49:31.6 & -00:46:30 & 6.9 & 0.28 &$ 17.3 \pm 0.1 $&$ 0.97 \pm 0.01 $&$ 1.16 \pm 0.01 $&$ 5.70 \pm 0.09 $&$ 0.10 \pm 0.00 $&$ 341 \pm 5 $&$ 378 \pm 5 $&$ 151 $&$ 0.90 \pm 0.02 $&$ 1.6 \pm 0.0 $ \\
\textbf{G31-10} & 18:49:31.8 & -00:38:03 & 6.9 & 0.56 &$ 20.5 \pm 0.6 $&$ 3.71 \pm 0.42 $&$ 13.78 \pm 1.56 $&$ 4.20 \pm 0.04 $&$ 0.10 \pm 0.00 $&$ 494 \pm 4 $&$ 3466 \pm 421 $&$ - $&$ 0.14 \pm 0.02 $&$ 14.4 \pm 1.8 $ \\
\textbf{G31-11} & 18:49:37.8 & -00:40:59 & 6.9 & 0.52 &$ 23.0 \pm 0.7 $&$ 2.10 \pm 0.13 $&$ 6.58 \pm 0.40 $&$ 2.41 \pm 0.04 $&$ 0.10 \pm 0.00 $&$ 260 \pm 5 $&$ 1406 \pm 103 $&$ - $&$ 0.19 \pm 0.01 $&$ 5.9 \pm 0.4 $ \\
\textbf{G31-12} & 18:49:34.6 & -00:38:21 & 6.9 & 0.44 &$ 21.6 \pm 1.0 $&$ 1.00 \pm 0.01 $&$ 2.28 \pm 0.02 $&$ 2.35 \pm 0.04 $&$ 0.10 \pm 0.01 $&$ 217 \pm 4 $&$ 533 \pm 34 $&$ - $&$ 0.41 \pm 0.03 $&$ 2.2 \pm 0.1 $ \\
\textbf{G31-13} & 18:49:37.6 & -00:39:55 & 6.9 & 0.40 &$ 19.9 \pm 0.2 $&$ 0.43 \pm 0.00 $&$ 0.79 \pm 0.00 $&$ 1.54 \pm 0.04 $&$ 0.10 \pm 0.01 $&$ 128 \pm 3 $&$ 207 \pm 3 $&$ - $&$ 0.62 \pm 0.02 $&$ 0.9 \pm 0.0 $ \\
G33-1 & 18:52:58.8 & +00:42:37 & 7.1 & 0.41 &$ 15.2 \pm 0.3 $&$ 2.03 \pm 0.07 $&$ 4.01 \pm 0.14 $&$ 3.07 \pm 0.05 $&$ 0.10 \pm 0.01 $&$ 264 \pm 5 $&$ 1708 \pm 85 $&$ 1135 $&$ 0.15 \pm 0.01 $&$ 6.5 \pm 0.3 $ \\
G33-4 & 18:52:56.4 & +00:43:08 & 7.1 & 0.59 &$ 15.2 \pm 0.3 $&$ 1.12 \pm 0.02 $&$ 4.15 \pm 0.08 $&$ 3.03 \pm 0.09 $&$ 0.37 \pm 0.06 $&$ 379 \pm 11 $&$ 1758 \pm 61 $&$ 820 $&$ 0.22 \pm 0.01 $&$ 6.7 \pm 0.2 $ \\
G33-7 & 18:52:58.1 & +00:44:08 & 7.1 & 0.79 &$ 17.7 \pm 0.9 $&$ 0.66 \pm 0.02 $&$ 3.95 \pm 0.10 $&$ 2.37 \pm 0.22 $&$ 0.11 \pm 0.22 $&$ 391 \pm 37 $&$ 1311 \pm 103 $&$ 641 $&$ 0.30 \pm 0.04 $&$ 5.0 \pm 0.4 $ \\
G33-8 & 18:52:53.9 & +00:41:16 & 7.1 & 0.73 &$ 17.7 \pm 0.5 $&$ 0.54 \pm 0.01 $&$ 2.85 \pm 0.07 $&$ 5.32 \pm 0.33 $&$ 0.10 \pm 0.05 $&$ 816 \pm 50 $&$ 946 \pm 44 $&$ 588 $&$ 0.86 \pm 0.07 $&$ 3.6 \pm 0.2 $ \\
G33-9 & 18:52:58.1 & +00:41:20 & 7.1 & 0.26 &$ 17.8 \pm 0.1 $&$ 0.46 \pm 0.00 $&$ 0.45 \pm 0.00 $&$ 2.15 \pm 0.17 $&$ 0.19 \pm 0.10 $&$ 117 \pm 10 $&$ 147 \pm 1 $&$ 119 $&$ 0.80 \pm 0.07 $&$ 0.6 \pm 0.0 $ \\
G33-11 & 18:52:56.2 & +00:41:48 & 7.1 & 0.30 &$ 17.2 \pm 0.0 $&$ 0.68 \pm 0.01 $&$ 0.81 \pm 0.01 $&$ 6.14 \pm 0.40 $&$ 0.10 \pm 0.01 $&$ 387 \pm 25 $&$ 282 \pm 4 $&$ 100 $&$ 1.37 \pm 0.09 $&$ 1.1 \pm 0.0 $ \\
G34-1 & 18:53:18.0 & +01:25:24 & 3.7 & 0.10 &$ 19.7 \pm 0.3 $&$ 11.29 \pm 0.74 $&$ 7.84 \pm 0.51 $&$ 2.88 \pm 0.00 $&$ 0.10 \pm 0.00 $&$ 57 \pm 0 $&$ 601 \pm 41 $&$ 1187 $&$ 0.10 \pm 0.01 $&$ 16.2 \pm 1.1 $ \\
G34-2 & 18:53:18.6 & +01:24:40 & 3.7 & 0.21 &$ 18.4 \pm 0.3 $&$ 7.30 \pm 0.72 $&$ 13.39 \pm 1.32 $&$ 3.43 \pm 0.01 $&$ 0.10 \pm 0.00 $&$ 151 \pm 1 $&$ 1139 \pm 116 $&$ 1284 $&$ 0.13 \pm 0.01 $&$ 30.8 \pm 3.1 $ \\
G34-3 & 18:53:20.4 & +01:28:23 & 3.7 & 0.19 &$ 16.3 \pm 0.1 $&$ 2.16 \pm 0.04 $&$ 3.38 \pm 0.07 $&$ 2.54 \pm 0.02 $&$ 0.10 \pm 0.00 $&$ 101 \pm 1 $&$ 347 \pm 8 $&$ 301 $&$ 0.29 \pm 0.01 $&$ 9.4 \pm 0.2 $ \\
G34-4 & 18:53:19.0 & +01:24:08 & 3.7 & 0.19 &$ 17.6 \pm 0.1 $&$ 4.50 \pm 0.48 $&$ 7.03 \pm 0.74 $&$ 2.70 \pm 0.02 $&$ 0.15 \pm 0.01 $&$ 108 \pm 1 $&$ 639 \pm 68 $&$ 253 $&$ 0.17 \pm 0.02 $&$ 17.3 \pm 1.8 $ \\
G34-5 & 18:53:19.8 & +01:23:30 & 3.7 & 0.45 &$ 17.4 \pm 0.3 $&$ 2.12 \pm 0.17 $&$ 14.93 \pm 1.18 $&$ 2.14 \pm 0.05 $&$ 0.13 \pm 0.03 $&$ 200 \pm 5 $&$ 1388 \pm 116 $&$ 664 $&$ 0.14 \pm 0.01 $&$ 37.5 \pm 3.1 $ \\
G34-6 & 18:53:18.6 & +01:27:48 & 3.7 & 0.31 &$ 16.6 \pm 0.2 $&$ 0.98 \pm 0.02 $&$ 3.62 \pm 0.06 $&$ 1.92 \pm 0.06 $&$ 0.12 \pm 0.04 $&$ 125 \pm 4 $&$ 362 \pm 8 $&$ 126 $&$ 0.34 \pm 0.01 $&$ 9.8 \pm 0.2 $ \\
G34-7 & 18:53:18.3 & +01:27:13 & 3.7 & 0.23 &$ 16.4 \pm 0.1 $&$ 0.76 \pm 0.00 $&$ 1.62 \pm 0.01 $&$ 2.21 \pm 0.09 $&$ 0.30 \pm 0.06 $&$ 107 \pm 4 $&$ 165 \pm 2 $&$ 87 $&$ 0.64 \pm 0.03 $&$ 4.5 \pm 0.0 $ \\
G34-8 & 18:53:16.4 & +01:26:20 & 3.7 & 0.26 &$ 17.7 \pm 0.6 $&$ 0.72 \pm 0.00 $&$ 1.88 \pm 0.01 $&$ 2.79 \pm 0.10 $&$ 0.12 \pm 0.05 $&$ 152 \pm 5 $&$ 170 \pm 10 $&$ 108 $&$ 0.90 \pm 0.06 $&$ 4.6 \pm 0.3 $ \\
G34-9 & 18:53:18.4 & +01:28:14 & 3.7 & 0.34 &$ 16.6 \pm 0.2 $&$ 1.49 \pm 0.05 $&$ 6.13 \pm 0.22 $&$ 2.30 \pm 0.04 $&$ 0.10 \pm 0.04 $&$ 162 \pm 2 $&$ 610 \pm 25 $&$ 157 $&$ 0.27 \pm 0.01 $&$ 16.5 \pm 0.7 $ \\
G35-4 & 18:57:06.7 & +02:08:23 & 2.9 & 0.23 &$ 15.7 \pm 0.8 $&$ 0.79 \pm 0.03 $&$ 2.47 \pm 0.08 $&$ 1.36 \pm 0.17 $&$ 0.10 \pm 0.05 $&$ 64 \pm 8 $&$ 167 \pm 14 $&$ 108 $&$ 0.39 \pm 0.06 $&$ 9.4 \pm 0.8 $ \\
G35-4 & 18:57:06.7 & +02:08:23 & 2.9 & 0.23 &$ 15.7 \pm 0.8 $&$ 0.79 \pm 0.03 $&$ 2.47 \pm 0.08 $&$ 1.15 \pm 0.06 $&$ 0.53 \pm 0.08 $&$ 54 \pm 3 $&$ 167 \pm 14 $&$ 108 $&$ 0.33 \pm 0.03 $&$ 9.4 \pm 0.8 $ \\
G35-5 & 18:57:08.8 & +02:08:09 & 2.9 & 0.24 &$ 14.6 \pm 0.3 $&$ 0.79 \pm 0.01 $&$ 2.95 \pm 0.02 $&$ 1.71 \pm 0.14 $&$ 0.10 \pm 0.03 $&$ 88 \pm 7 $&$ 224 \pm 9 $&$ 118 $&$ 0.39 \pm 0.03 $&$ 12.6 \pm 0.5 $ \\
G35-5 & 18:57:08.8 & +02:08:09 & 2.9 & 0.24 &$ 14.6 \pm 0.3 $&$ 0.79 \pm 0.01 $&$ 2.95 \pm 0.02 $&$ 1.57 \pm 0.07 $&$ 0.37 \pm 0.04 $&$ 81 \pm 4 $&$ 224 \pm 9 $&$ 118 $&$ 0.36 \pm 0.02 $&$ 12.6 \pm 0.5 $ \\
G35-6 & 18:57:08.4 & +02:09:09 & 2.9 & 0.24 &$ 16.0 \pm 0.2 $&$ 0.77 \pm 0.01 $&$ 2.86 \pm 0.03 $&$ 1.27 \pm 0.05 $&$ 0.60 \pm 0.08 $&$ 65 \pm 3 $&$ 186 \pm 4 $&$ 71 $&$ 0.35 \pm 0.02 $&$ 10.4 \pm 0.3 $ \\
G35-7 & 18:57:08.1 & +02:10:50 & 2.9 & 0.26 &$ 14.8 \pm 0.3 $&$ 0.92 \pm 0.01 $&$ 3.78 \pm 0.05 $&$ 1.85 \pm 0.04 $&$ 0.36 \pm 0.03 $&$ 101 \pm 2 $&$ 281 \pm 10 $&$ 96 $&$ 0.36 \pm 0.02 $&$ 15.8 \pm 0.6 $ \\
G35-8 & 18:57:07.0 & +02:08:54 & 2.9 & 0.18 &$ 16.1 \pm 0.5 $&$ 0.59 \pm 0.00 $&$ 1.36 \pm 0.01 $&$ 1.52 \pm 0.05 $&$ 0.27 \pm 0.05 $&$ 59 \pm 2 $&$ 87 \pm 4 $&$ 59 $&$ 0.68 \pm 0.04 $&$ 4.9 \pm 0.2 $ \\
G35-9 & 18:57:11.2 & +02:07:27 & 2.9 & 0.17 &$ 14.0 \pm 0.3 $&$ 0.72 \pm 0.01 $&$ 1.43 \pm 0.02 $&$ 1.39 \pm 0.05 $&$ 0.73 \pm 0.09 $&$ 50 \pm 2 $&$ 115 \pm 4 $&$ 42 $&$ 0.43 \pm 0.02 $&$ 6.5 \pm 0.2 $ \\
\textbf{G35-10} & 18:57:07.7 & +02:11:50 & 2.9 & 0.20 &$ 15.1 \pm 0.3 $&$ 0.48 \pm 0.01 $&$ 1.16 \pm 0.02 $&$ 1.23 \pm 0.09 $&$ 0.26 \pm 0.07 $&$ 52 \pm 4 $&$ 83 \pm 3 $&$ - $&$ 0.62 \pm 0.05 $&$ 4.6 \pm 0.2 $ \\
\textbf{G35-11} & 18:57:05.9 & +02:12:30 & 2.9 & 0.13 &$ 16.7 \pm 0.1 $&$ 0.28 \pm 0.00 $&$ 0.31 \pm 0.00 $&$ 0.78 \pm 0.05 $&$ 0.90 \pm 0.18 $&$ 21 \pm 1 $&$ 19 \pm 0 $&$ - $&$ 1.14 \pm 0.08 $&$ 1.0 \pm 0.0 $ \\
G38-1 & 19:04:07.4 & +05:08:48 & 2.7 & 0.17 &$ 15.2 \pm 0.1 $&$ 1.33 \pm 0.03 $&$ 3.05 \pm 0.06 $&$ 1.94 \pm 0.06 $&$ 0.33 \pm 0.05 $&$ 71 \pm 2 $&$ 186 \pm 5 $&$ 117 $&$ 0.38 \pm 0.01 $&$ 13.0 \pm 0.3 $ \\
G38-2 & 19:04:03.4 & +05:07:56 & 2.7 & 0.14 &$ 18.5 \pm 0.3 $&$ 1.18 \pm 0.02 $&$ 2.01 \pm 0.03 $&$ 1.27 \pm 0.06 $&$ 0.15 \pm 0.07 $&$ 37 \pm 2 $&$ 91 \pm 2 $&$ 73 $&$ 0.41 \pm 0.02 $&$ 6.3 \pm 0.2 $ \\
G38-3 & 19:04:07.4 & +05:09:44 & 2.7 & 0.07 &$ 16.2 \pm 0.0 $&$ 0.93 \pm 0.00 $&$ 0.36 \pm 0.00 $&$ 1.72 \pm 0.10 $&$ 0.30 \pm 0.08 $&$ 25 \pm 1 $&$ 20 \pm 0 $&$ 11 $&$ 1.26 \pm 0.07 $&$ 1.4 \pm 0.0 $ \\
G38-4 & 19:04:00.6 & +05:09:06 & 2.7 & 0.20 &$ 18.1 \pm 0.3 $&$ 0.66 \pm 0.01 $&$ 1.94 \pm 0.02 $&$ 1.27 \pm 0.12 $&$ 0.22 \pm 0.14 $&$ 53 \pm 5 $&$ 90 \pm 2 $&$ 48 $&$ 0.59 \pm 0.06 $&$ 6.3 \pm 0.2 $ \\
\textbf{G38-5} & 19:04:07.4 & +05:07:08 & 2.7 & 0.18 &$ 19.2 \pm 0.1 $&$ 0.31 \pm 0.00 $&$ 0.76 \pm 0.01 $&$ 0.98 \pm 0.08 $&$ 0.10 \pm 0.39 $&$ 37 \pm 3 $&$ 32 \pm 1 $&$ - $&$ 1.15 \pm 0.10 $&$ 2.2 \pm 0.0 $ \\
G53-1 & 19:29:17.2 & +17:56:21 & 1.8 & 0.07 &$ 19.3 \pm 0.2 $&$ 4.55 \pm 0.60 $&$ 5.44 \pm 0.72 $&$ 1.83 \pm 0.03 $&$ 0.10 \pm 0.01 $&$ 29 \pm 0 $&$ 102 \pm 14 $&$ 124 $&$ 0.28 \pm 0.04 $&$ 24.0 \pm 3.2 $ \\
G53-2 & 19:29:20.2 & +17:57:06 & 1.8 & 0.14 &$ 17.3 \pm 1.0 $&$ 0.77 \pm 0.02 $&$ 2.27 \pm 0.06 $&$ 1.23 \pm 0.03 $&$ 0.17 \pm 0.03 $&$ 35 \pm 1 $&$ 50 \pm 5 $&$ 44 $&$ 0.69 \pm 0.07 $&$ 11.8 \pm 1.1 $ \\
G53-4 & 19:29:20.4 & +17:55:04 & 1.8 & 0.18 &$ 17.2 \pm 0.3 $&$ 0.33 \pm 0.01 $&$ 1.72 \pm 0.03 $&$ 0.59 \pm 0.04 $&$ 0.71 \pm 0.20 $&$ 23 \pm 1 $&$ 39 \pm 1 $&$ 45 $&$ 0.60 \pm 0.04 $&$ 9.1 \pm 0.3 $ \\
\hline
\end{tabular}
\begin{itemize}
\item The distances (in Column 4) of these IRDCs refer to the works from \citet{Rathborne2006} and \citet{Simon2006}. $R_{\rm core}$ (in Column 5) indicates the radius of core. The duplicated sources, for example, G31-3, indicate that the corresponding spectra have two velocity components within core size.
\end{itemize}
\end{table*}
\end{landscape}

\begin{landscape}
\begin{table*}
\caption{Gaussian fitting parameters (integrated intensity, velocity, FWHM, and intensity of $\hcop$, HNC, and $\co$) with masking absorption dip toward the selected cores in the IRDCs.}
\label{tab:hcop} \centering 
\footnotesize
\begin{tabular}{lcccccccccccc}
\hline \hline
Core & $S_{\hcop}$ & $V_{\hcop}$ & $\Delta V_{\hcop}$ & $I_{\hcop}$ & $S_{\rm HNC}$ & $V_{\rm HNC}$ & $\Delta V_{\rm HNC}$ & $I_{\rm HNC}$ & $S_{\co}$ & $V_{\co}$ & $\Delta V_{\co}$ & $I_{\co}$ \\
 & K km s$^{-1}$ & km s$^{-1}$ & km s$^{-1}$ & K & K km s$^{-1}$ & km s$^{-1}$ & km s$^{-1}$ & K & K km s$^{-1}$ & km s$^{-1}$ & km s$^{-1}$ & K \\
\hline
G31-1   &$      47.84   \pm     0.57    $&$     95.25   \pm     0.03    $&$     7.13    \pm     0.05    $&$     6.31    \pm     0.08    $&$     30.30   \pm     0.39    $&$     95.04   \pm     0.03    $&$     5.78    \pm     0.05    $&$     4.92    \pm     0.08    $&$     16.19   \pm     0.65    $&$     95.79   \pm     0.04    $&$     3.80    \pm     0.11    $&$     4.01    \pm     0.18    $\\
G31-2   &$      52.60   \pm     0.73    $&$     94.65   \pm     0.03    $&$     6.37    \pm     0.04    $&$     7.76    \pm     0.08    $&$     27.31   \pm     0.39    $&$     94.44   \pm     0.03    $&$     5.73    \pm     0.05    $&$     4.48    \pm     0.08    $&$     18.21   \pm     1.11    $&$     95.41   \pm     0.04    $&$     5.18    \pm     0.15    $&$     3.30    \pm     0.11    $\\
G31-3   &$      16.21   \pm     0.16    $&$     93.67   \pm     0.02    $&$     4.12    \pm     0.05    $&$     3.69    \pm     0.09    $&$     12.43   \pm     0.17    $&$     93.89   \pm     0.02    $&$     3.53    \pm     0.05    $&$     3.31    \pm     0.08    $&$     3.79    \pm     0.01    $&$     94.59   \pm     0.11    $&$     4.69    \pm     0.06    $&$     0.76    \pm     0.12    $\\
G31-3   &$      13.28   \pm     0.16    $&$     99.68   \pm     0.02    $&$     4.30    \pm     0.06    $&$     2.90    \pm     0.09    $&$     12.27   \pm     0.17    $&$     99.29   \pm     0.03    $&$     4.57    \pm     0.07    $&$     2.52    \pm     0.08    $&$     9.49    \pm     0.14    $&$     98.72   \pm     0.03    $&$     3.59    \pm     0.04    $&$     2.48    \pm     0.12    $\\
G31-4   &$      34.69   \pm     0.61    $&$     96.60   \pm     0.02    $&$     6.22    \pm     0.07    $&$     5.24    \pm     0.08    $&$     31.29   \pm     0.71    $&$     96.81   \pm     0.04    $&$     5.05    \pm     0.05    $&$     5.82    \pm     0.09    $&$     14.74   \pm     3.03    $&$     97.10   \pm     0.07    $&$     3.81    \pm     0.25    $&$     3.64    \pm     0.12    $\\
G31-5   &$      12.44   \pm     0.83    $&$     96.22   \pm     0.05    $&$     5.46    \pm     0.21    $&$     2.14    \pm     0.08    $&$     6.83    \pm     0.14    $&$     96.42   \pm     0.05    $&$     4.79    \pm     0.10    $&$     1.34    \pm     0.09    $&$     4.91    \pm     0.13    $&$     96.62   \pm     0.04    $&$     3.13    \pm     0.10    $&$     1.47    \pm     0.11    $\\
G31-6   &$      45.70   \pm     0.35    $&$     95.15   \pm     0.02    $&$     6.55    \pm     0.03    $&$     6.55    \pm     0.08    $&$     33.09   \pm     0.47    $&$     95.26   \pm     0.04    $&$     5.81    \pm     0.05    $&$     5.35    \pm     0.09    $&$     26.85   \pm     0.84    $&$     96.56   \pm     0.02    $&$     4.47    \pm     0.08    $&$     5.65    \pm     0.12    $\\
G31-7   &$      16.56   \pm     0.60    $&$     94.85   \pm     0.08    $&$     5.32    \pm     0.09    $&$     2.92    \pm     0.08    $&$     5.86    \pm     0.09    $&$     93.50   \pm     0.02    $&$     3.13    \pm     0.06    $&$     1.76    \pm     0.09    $&$     9.80    \pm     0.82    $&$     95.47   \pm     0.08    $&$     4.08    \pm     0.18    $&$     2.26    \pm     0.13    $\\
G31-8   &$      19.26   \pm     0.36    $&$     95.11   \pm     0.04    $&$     7.60    \pm     0.11    $&$     2.38    \pm     0.08    $&$     14.85   \pm     0.41    $&$     95.37   \pm     0.05    $&$     5.41    \pm     0.09    $&$     2.58    \pm     0.08    $&$     10.17   \pm     0.46    $&$     96.48   \pm     0.07    $&$     4.24    \pm     0.13    $&$     2.25    \pm     0.11    $\\
G31-9   &$      25.45   \pm     0.26    $&$     95.83   \pm     0.03    $&$     7.24    \pm     0.07    $&$     3.30    \pm     0.08    $&$     15.43   \pm     0.15    $&$     96.38   \pm     0.03    $&$     6.30    \pm     0.08    $&$     2.30    \pm     0.08    $&$     8.17    \pm     0.13    $&$     97.55   \pm     0.03    $&$     3.81    \pm     0.08    $&$     2.01    \pm     0.12    $\\
\textbf{G31-10} &$      27.12   \pm     0.28    $&$     93.39   \pm     0.02    $&$     5.00    \pm     0.05    $&$     5.10    \pm     0.10    $&$     20.50   \pm     0.28    $&$     93.71   \pm     0.02    $&$     4.69    \pm     0.06    $&$     4.11    \pm     0.11    $&$     14.77   \pm     0.32    $&$     93.64   \pm     0.03    $&$     3.79    \pm     0.09    $&$     3.66    \pm     0.20    $\\
\textbf{G31-11} &$      16.71   \pm     0.23    $&$     96.36   \pm     0.03    $&$     5.18    \pm     0.06    $&$     3.03    \pm     0.09    $&$     13.29   \pm     0.18    $&$     96.39   \pm     0.03    $&$     4.21    \pm     0.07    $&$     2.97    \pm     0.10    $&$     12.91   \pm     0.17    $&$     96.22   \pm     0.02    $&$     3.01    \pm     0.05    $&$     4.02    \pm     0.17    $\\
\textbf{G31-12} &$      11.59   \pm     0.29    $&$     94.72   \pm     0.04    $&$     4.18    \pm     0.07    $&$     2.61    \pm     0.10    $&$     6.85    \pm     0.11    $&$     95.72   \pm     0.02    $&$     2.73    \pm     0.06    $&$     2.36    \pm     0.11    $&$     6.21    \pm     0.31    $&$     95.29   \pm     0.04    $&$     2.81    \pm     0.17    $&$     2.08    \pm     0.19    $\\
\textbf{G31-13} &$      2.62    \pm     0.10    $&$     97.03   \pm     0.04    $&$     2.64    \pm     0.13    $&$     0.93    \pm     0.09    $&$     3.06    \pm     0.09    $&$     96.51   \pm     0.03    $&$     2.03    \pm     0.08    $&$     1.42    \pm     0.11    $&$     4.43    \pm     0.15    $&$     96.54   \pm     0.03    $&$     2.03    \pm     0.08    $&$     2.05    \pm     0.18    $\\
G33-1   &$      25.35   \pm     1.12    $&$     106.40  \pm     0.06    $&$     5.88    \pm     0.16    $&$     4.05    \pm     0.15    $&$     20.13   \pm     0.45    $&$     106.00  \pm     0.05    $&$     4.69    \pm     0.08    $&$     4.03    \pm     0.16    $&$     6.20    \pm     0.25    $&$     106.60  \pm     0.07    $&$     3.69    \pm     0.18    $&$     1.58    \pm     0.23    $\\
G33-4   &$      12.90   \pm     0.61    $&$     106.30  \pm     0.15    $&$     7.20    \pm     0.24    $&$     1.68    \pm     0.15    $&$     10.82   \pm     0.44    $&$     106.20  \pm     0.07    $&$     4.97    \pm     0.18    $&$     2.05    \pm     0.16    $&$     6.50    \pm     0.27    $&$     106.40  \pm     0.09    $&$     4.39    \pm     0.21    $&$     1.39    \pm     0.23    $\\
G33-7   &$      4.83    \pm     0.29    $&$     107.00  \pm     0.16    $&$     5.55    \pm     0.36    $&$     0.82    \pm     0.15    $&$     4.27    \pm     0.32    $&$     106.50  \pm     0.14    $&$     4.19    \pm     0.28    $&$     0.96    \pm     0.15    $&$     6.80    \pm     0.69    $&$     106.30  \pm     0.10    $&$     2.62    \pm     0.14    $&$     2.44    \pm     0.23    $\\
G33-8   &$      14.03   \pm     0.43    $&$     104.70  \pm     0.12    $&$     7.58    \pm     0.19    $&$     1.74    \pm     0.14    $&$     9.55    \pm     0.62    $&$     103.80  \pm     0.23    $&$     6.75    \pm     0.32    $&$     1.33    \pm     0.15    $&$     11.29   \pm     1.04    $&$     104.60  \pm     0.11    $&$     6.41    \pm     0.42    $&$     1.66    \pm     0.22    $\\
G33-9   &$      7.13    \pm     0.33    $&$     105.40  \pm     0.16    $&$     7.16    \pm     0.31    $&$     0.94    \pm     0.14    $&$     4.38    \pm     0.84    $&$     106.10  \pm     0.28    $&$     3.57    \pm     0.34    $&$     1.15    \pm     0.15    $&$     3.83    \pm     0.23    $&$     106.70  \pm     0.09    $&$     3.22    \pm     0.23    $&$     1.12    \pm     0.23    $\\
G33-11  &$      14.93   \pm     0.45    $&$     105.20  \pm     0.13    $&$     8.37    \pm     0.20    $&$     1.67    \pm     0.15    $&$     9.58    \pm     0.27    $&$     104.70  \pm     0.10    $&$     7.45    \pm     0.24    $&$     1.21    \pm     0.16    $&$     10.15   \pm     0.64    $&$     105.30  \pm     0.16    $&$     6.81    \pm     0.35    $&$     1.40    \pm     0.23    $\\
G34-1   &$      19.42   \pm     0.23    $&$     58.85   \pm     0.03    $&$     5.28    \pm     0.06    $&$     3.45    \pm     0.09    $&$     26.62   \pm     0.15    $&$     58.07   \pm     0.01    $&$     4.48    \pm     0.03    $&$     5.58    \pm     0.10    $&$     8.21    \pm     0.19    $&$     57.31   \pm     0.04    $&$     3.41    \pm     0.09    $&$     2.27    \pm     0.15    $\\
G34-2   &$      29.70   \pm     1.25    $&$     57.82   \pm     0.11    $&$     9.89    \pm     0.21    $&$     2.82    \pm     0.10    $&$     27.25   \pm     0.31    $&$     57.81   \pm     0.03    $&$     5.76    \pm     0.05    $&$     4.44    \pm     0.11    $&$     20.89   \pm     0.22    $&$     56.98   \pm     0.02    $&$     4.12    \pm     0.04    $&$     4.76    \pm     0.15    $\\
G34-3   &$      17.99   \pm     0.23    $&$     60.81   \pm     0.02    $&$     3.78    \pm     0.05    $&$     4.47    \pm     0.12    $&$     14.87   \pm     0.14    $&$     59.70   \pm     0.02    $&$     4.29    \pm     0.05    $&$     3.26    \pm     0.11    $&$     4.89    \pm     0.12    $&$     58.67   \pm     0.03    $&$     2.39    \pm     0.07    $&$     1.92    \pm     0.14    $\\
G34-4   &$      14.02   \pm     0.52    $&$     57.35   \pm     0.17    $&$     12.87   \pm     0.46    $&$     1.02    \pm     0.11    $&$     9.87    \pm     0.15    $&$     56.65   \pm     0.02    $&$     3.18    \pm     0.06    $&$     2.92    \pm     0.11    $&$     13.99   \pm     0.31    $&$     57.52   \pm     0.03    $&$     3.70    \pm     0.07    $&$     3.55    \pm     0.16    $\\
G34-5   &$      3.50    \pm     0.18    $&$     57.89   \pm     0.10    $&$     4.82    \pm     0.30    $&$     0.68    \pm     0.10    $&$     6.84    \pm     0.14    $&$     57.88   \pm     0.03    $&$     3.07    \pm     0.07    $&$     2.09    \pm     0.10    $&$     6.41    \pm     0.12    $&$     57.73   \pm     0.02    $&$     2.28    \pm     0.05    $&$     2.64    \pm     0.14    $\\
G34-6   &$      3.76    \pm     0.11    $&$     59.50   \pm     0.03    $&$     2.22    \pm     0.09    $&$     1.59    \pm     0.11    $&$     6.55    \pm     0.12    $&$     58.61   \pm     0.03    $&$     3.18    \pm     0.07    $&$     1.94    \pm     0.11    $&$     6.39    \pm     0.13    $&$     58.17   \pm     0.02    $&$     2.24    \pm     0.05    $&$     2.69    \pm     0.17    $\\
G34-7   &$      1.74    \pm     0.11    $&$     59.48   \pm     0.07    $&$     2.41    \pm     0.20    $&$     0.68    \pm     0.10    $&$     4.54    \pm     0.15    $&$     58.21   \pm     0.04    $&$     3.26    \pm     0.12    $&$     1.31    \pm     0.10    $&$     7.25    \pm     0.13    $&$     57.90   \pm     0.02    $&$     2.36    \pm     0.05    $&$     2.88    \pm     0.15    $\\
G34-8   &$      1.73    \pm     0.08    $&$     59.15   \pm     0.04    $&$     1.83    \pm     0.11    $&$     0.89    \pm     0.10    $&$     5.42    \pm     0.20    $&$     57.59   \pm     0.04    $&$     3.32    \pm     0.11    $&$     1.53    \pm     0.10    $&$     11.44   \pm     0.39    $&$     57.37   \pm     0.03    $&$     2.73    \pm     0.09    $&$     3.93    \pm     0.24    $\\
G34-9   &$      5.68    \pm     0.11    $&$     59.61   \pm     0.02    $&$     2.69    \pm     0.06    $&$     1.98    \pm     0.10    $&$     8.91    \pm     0.12    $&$     58.69   \pm     0.02    $&$     3.52    \pm     0.05    $&$     2.38    \pm     0.10    $&$     5.61    \pm     0.12    $&$     57.96   \pm     0.02    $&$     1.89    \pm     0.05    $&$     2.78    \pm     0.15    $\\
G35-4   &$      8.93    \pm     0.12    $&$     43.76   \pm     0.02    $&$     2.90    \pm     0.05    $&$     2.89    \pm     0.10    $&$     8.17    \pm     0.12    $&$     44.15   \pm     0.02    $&$     3.04    \pm     0.05    $&$     2.53    \pm     0.10    $&$     5.41    \pm     0.19    $&$     45.68   \pm     0.06    $&$     2.84    \pm     0.08    $&$     1.79    \pm     0.13    $\\
G35-5   &$      9.05    \pm     0.12    $&$     43.85   \pm     0.03    $&$     4.13    \pm     0.07    $&$     2.06    \pm     0.09    $&$     9.66    \pm     0.10    $&$     44.29   \pm     0.02    $&$     3.71    \pm     0.04    $&$     2.45    \pm     0.09    $&$     3.32    \pm     0.10    $&$     45.93   \pm     0.03    $&$     2.27    \pm     0.09    $&$     1.38    \pm     0.12    $\\
G35-6   &$      7.72    \pm     0.17    $&$     44.94   \pm     0.05    $&$     4.72    \pm     0.12    $&$     1.54    \pm     0.09    $&$     6.17    \pm     0.14    $&$     45.35   \pm     0.02    $&$     2.40    \pm     0.06    $&$     2.42    \pm     0.08    $&$     7.82    \pm     0.32    $&$     44.39   \pm     0.03    $&$     2.44    \pm     0.08    $&$     3.01    \pm     0.14    $\\
G35-7   &$      11.81   \pm     0.15    $&$     45.77   \pm     0.02    $&$     4.01    \pm     0.06    $&$     2.77    \pm     0.09    $&$     8.52    \pm     0.14    $&$     45.79   \pm     0.02    $&$     2.71    \pm     0.04    $&$     2.95    \pm     0.09    $&$     7.34    \pm     0.38    $&$     44.55   \pm     0.04    $&$     2.47    \pm     0.09    $&$     2.79    \pm     0.15    $\\
G35-8   &$      8.33    \pm     0.16    $&$     44.59   \pm     0.04    $&$     4.15    \pm     0.08    $&$     1.89    \pm     0.09    $&$     5.81    \pm     0.19    $&$     44.95   \pm     0.04    $&$     2.59    \pm     0.06    $&$     2.11    \pm     0.09    $&$     3.37    \pm     0.09    $&$     45.26   \pm     0.02    $&$     1.57    \pm     0.05    $&$     2.01    \pm     0.13    $\\
G35-9   &$      7.69    \pm     0.17    $&$     44.69   \pm     0.02    $&$     2.47    \pm     0.06    $&$     2.92    \pm     0.09    $&$     5.45    \pm     0.11    $&$     44.87   \pm     0.02    $&$     2.46    \pm     0.05    $&$     2.08    \pm     0.09    $&$     5.85    \pm     0.24    $&$     45.69   \pm     0.08    $&$     2.85    \pm     0.15    $&$     1.93    \pm     0.13    $\\
\textbf{G35-10} &$      6.14    \pm     0.11    $&$     45.88   \pm     0.03    $&$     3.94    \pm     0.08    $&$     1.46    \pm     0.09    $&$     4.69    \pm     0.09    $&$     45.96   \pm     0.02    $&$     2.54    \pm     0.06    $&$     1.73    \pm     0.09    $&$     4.01    \pm     0.31    $&$     44.94   \pm     0.09    $&$     2.15    \pm     0.15    $&$     1.75    \pm     0.14    $\\
\textbf{G35-11} &$      4.96    \pm     0.09    $&$     45.99   \pm     0.03    $&$     3.26    \pm     0.07    $&$     1.43    \pm     0.09    $&$     3.02    \pm     0.08    $&$     46.12   \pm     0.03    $&$     2.27    \pm     0.07    $&$     1.25    \pm     0.09    $&$     5.32    \pm     0.08    $&$     44.90   \pm     0.01    $&$     1.45    \pm     0.02    $&$     3.46    \pm     0.12    $\\
G38-1   &$      14.86   \pm     0.21    $&$     42.01   \pm     0.02    $&$     3.79    \pm     0.06    $&$     3.69    \pm     0.13    $&$     11.54   \pm     0.21    $&$     42.16   \pm     0.02    $&$     2.81    \pm     0.05    $&$     3.86    \pm     0.12    $&$     13.28   \pm     0.20    $&$     42.07   \pm     0.02    $&$     2.35    \pm     0.03    $&$     5.32    \pm     0.17    $\\
G38-2   &$      8.06    \pm     0.13    $&$     41.63   \pm     0.02    $&$     3.12    \pm     0.06    $&$     2.43    \pm     0.11    $&$     5.50    \pm     0.09    $&$     41.82   \pm     0.02    $&$     2.34    \pm     0.04    $&$     2.21    \pm     0.10    $&$     8.20    \pm     0.09    $&$     41.63   \pm     0.01    $&$     1.38    \pm     0.02    $&$     5.57    \pm     0.14    $\\
G38-3   &$      12.19   \pm     0.22    $&$     42.57   \pm     0.04    $&$     4.44    \pm     0.08    $&$     2.58    \pm     0.12    $&$     6.07    \pm     0.13    $&$     42.92   \pm     0.03    $&$     2.93    \pm     0.08    $&$     1.95    \pm     0.12    $&$     6.44    \pm     0.16    $&$     42.09   \pm     0.03    $&$     2.77    \pm     0.08    $&$     2.18    \pm     0.17    $\\
G38-4   &$      5.83    \pm     0.14    $&$     42.06   \pm     0.04    $&$     3.11    \pm     0.09    $&$     1.76    \pm     0.12    $&$     3.97    \pm     0.11    $&$     42.24   \pm     0.03    $&$     2.07    \pm     0.06    $&$     1.80    \pm     0.12    $&$     6.08    \pm     0.13    $&$     42.00   \pm     0.02    $&$     1.61    \pm     0.04    $&$     3.54    \pm     0.18    $\\
\textbf{G38-5}  &$      6.30    \pm     0.10    $&$     41.51   \pm     0.01    $&$     1.86    \pm     0.04    $&$     3.19    \pm     0.11    $&$     4.54    \pm     0.08    $&$     41.60   \pm     0.01    $&$     1.58    \pm     0.03    $&$     2.71    \pm     0.10    $&$     3.29    \pm     0.10    $&$     41.57   \pm     0.02    $&$     1.50    \pm     0.06    $&$     2.05    \pm     0.15    $\\
G53-1   &$      34.47   \pm     0.26    $&$     21.75   \pm     0.02    $&$     4.00    \pm     0.03    $&$     8.09    \pm     0.13    $&$     14.32   \pm     0.11    $&$     21.55   \pm     0.01    $&$     2.63    \pm     0.03    $&$     5.11    \pm     0.10    $&$     8.80    \pm     0.12    $&$     21.85   \pm     0.02    $&$     2.59    \pm     0.04    $&$     3.20    \pm     0.14    $\\
G53-2   &$      13.89   \pm     0.31    $&$     23.05   \pm     0.03    $&$     3.37    \pm     0.06    $&$     3.87    \pm     0.13    $&$     7.78    \pm     0.11    $&$     22.68   \pm     0.01    $&$     1.89    \pm     0.03    $&$     3.86    \pm     0.12    $&$     5.00    \pm     0.13    $&$     22.49   \pm     0.03    $&$     2.01    \pm     0.06    $&$     2.34    \pm     0.17    $\\
G53-4   &$      4.22    \pm     0.12    $&$     21.64   \pm     0.03    $&$     2.33    \pm     0.08    $&$     1.70    \pm     0.12    $&$     3.64    \pm     0.09    $&$     21.64   \pm     0.02    $&$     1.32    \pm     0.04    $&$     2.58    \pm     0.12    $&$     3.51    \pm     0.10    $&$     21.68   \pm     0.02    $&$     1.20    \pm     0.04    $&$     2.74    \pm     0.17    $\\
\hline
\end{tabular}
\begin{itemize}
\item The duplicated sources, for example, G31-3, indicate that the corresponding spectra have two velocity components within core size.
\end{itemize}
\end{table*}
\end{landscape}

\begin{table*}
\caption{Parameters of the detected DCO$^{+}$, SiO, and DCN in the IRDCs.}
\label{tab_cso-spectra} \centering 
\begin{tabular}{cccccccc}
\hline \hline
Core & Velocity & $\Delta\, V$(DCO$^{+}$) & $T_{\rm MB}$(DCO$^{+}$)  & $\Delta\, V$(SiO)  & $T_{\rm MB}$(SiO)   &
$\Delta\, V$(DCN) & $T_{\rm MB}$(DCN) \\
         &  $\kms$  & $\kms$ & K  & $\kms$ & K   & $\kms$ & K    \\
 \hline
 G31-1  & $  95.53  \pm  0.56 $  &  $   3.40    \pm  1.51   $  &  $   0.050   \pm   0.028    $  &  $  11.06    \pm  0.56
  $  &  $   0.209   \pm   0.030    $  &  $   6.23    \pm  0.63   $  &  $   0.106   \pm   0.026    $   \\
 G34-1  & $  58.84  \pm  0.13 $  &  $   1.88    \pm  0.39   $  &  $   0.183   \pm   0.037    $  &  $  10.50    \pm  0.45
  $  &  $   0.343   \pm   0.040    $  &  $   4.47    \pm  0.34   $  &  $   0.306   \pm   0.036    $   \\
 G34-2  & $  57.47  \pm  0.14 $  &  $   3.24    \pm  0.32   $  &  $   0.193   \pm   0.033    $  &  $   9.46    \pm  0.51
  $  &  $   0.308   \pm   0.038    $  &  $   4.99    \pm  0.37   $  &  $   0.195   \pm   0.032    $   \\
 G34-3  & $  59.92  \pm  0.15 $  &  $   1.58    \pm  0.38   $  &  $   0.120   \pm   0.031    $  &  $  17.49    \pm  1.32
  $  &  $   0.160   \pm   0.037    $  &  $  11.68    \pm  2.95   $  &  $   0.038   \pm   0.030    $   \\
 G38-1  & $  42.91  \pm  0.07 $  &  $   1.92    \pm  0.18   $  &  $   0.366   \pm   0.041    $  &  $   5.01    \pm  1.29
  $  &  $   0.075   \pm   0.043    $  &  $   -   $  &  $   -    $   \\
 G53-1  & $  22.60  \pm  0.17 $  &  $   2.02    \pm  0.30   $  &  $   0.406   \pm   0.113    $  &  $  12.54    \pm  1.37
  $  &  $   0.171   \pm   0.064    $  &  $   5.73    \pm  0.83   $  &  $   0.181   \pm   0.052    $   \\
  \hline
\end{tabular}
\end{table*}

\end{document}